%% file: DMACS_ieee.tex
\documentclass[10pt, conference, letterpaper]{IEEEtran}

\usepackage{cite}

\usepackage{subfigure}

\usepackage{url}
\usepackage{verbatim,epsf}
\usepackage[caption=false,font=footnotesize]{subfig}

\usepackage{tabularx}
\usepackage{balance} 

\usepackage{amssymb}
\usepackage{amsmath}
\usepackage{graphicx}
\usepackage{epstopdf}

\usepackage{epsfig, color}

\usepackage{comment} 
\usepackage{amsfonts} 
\usepackage{amsmath} 
\usepackage{rotating}

\usepackage{amssymb}
\usepackage{amsmath}
\usepackage{amsfonts}

\usepackage{verbatim}

\newtheorem{theorem}{Theorem}
\newtheorem{lemma}{Lemma}

\newtheorem{definition}{Definition}

\usepackage{algorithm}
\usepackage{algorithmic}

\newsavebox{\ieeealgbox}

\begin{document}

\title{Auction-based Incentive Mechanisms for\\Dynamic Mobile Ad-Hoc Crowd Service }

\author{\IEEEauthorblockN{Honggang Zhang\IEEEauthorrefmark{1}, 
Benyuan Liu\IEEEauthorrefmark{2}, Hengky Susanto\IEEEauthorrefmark{2}, 
Guoliang Xue\IEEEauthorrefmark{3}\\
\IEEEauthorrefmark{1} CIS Dept., Fordham University, Bronx, NY.  Email: honggang@cis.fordham.edu \\
\IEEEauthorrefmark{2} CS Dept., UMass Lowell, Lowell, MA. Email: \{bliu,hsusanto\}@cs.uml.edu}
\IEEEauthorrefmark{3} School of CIDSE, Arizona State University, Tempe,   AZ.  Email: xue@asu.edu
}

\maketitle

\begin{abstract}

We investigate a type of emerging user-assisted mobile applications or services, 
referred to as Dynamic Mobile Ad-hoc Crowd Service (DMACS), such as
collaborative streaming via smartphones 
or location privacy protection through a crowd of smartphone users.
Such services are provided and consumed by users carrying smart mobile devices (e.g., smartphones)
who are in close proximity of each other (e.g., within Bluetooth range). 
Users in a DMACS system dynamically arrive and depart over time, 
and are divided into multiple possibly overlapping groups according to radio range constraints. 
Crucial to the success of such systems is a mechanism that
incentivizes users' participation and ensures fair trading. 
In this paper, we design a multi-market, dynamic double auction mechanism, referred to as 
M-CHAIN, and show that it is 
truthful, feasible, individual-rational, no-deficit, and computationally efficient.
The novelty and significance of M-CHAIN is that  
\textcolor{black}{it addresses and solves the fair trading problem}
in a multi-group or multi-market dynamic double auction problem 
which naturally occurs in a mobile wireless environment.
We demonstrate its efficiency via 
simulations based on generated user patterns
(stochastic arrivals, random market clustering of users)
and real-world traces.
\end{abstract}

\section{Introduction}\label{sec:introduction}

Recent advances in 
smart mobile devices (e.g., smartphones) 
have enabled many powerful mobile applications and services that
are not available on desktops or laptops, 
thanks to these devices' embedded sensors (e.g., GPS), 
increasingly powerful processors, advanced networking capabilities (e.g., Bluetooth, Wi-Fi),
and fast cellular network access.
In this paper we consider a type of emerging user-assisted mobile applications or services, 
referred to as \textit{Dynamic Mobile Ad-hoc Crowd Service (DMACS)}, and
focus on designing incentive mechanisms that are crucial to the success of such services.
An example of DMACS is cooperative streaming \cite{keller2012microcast} (a data sharing service) 
in which users cooperatively stream to their phones a high quality video using
their data plans\footnote{Due to bandwidth limitation and data plan quota, 
only low quality videos can be streamed to a single phone through a single user's data plan.}. 
Another example is a Peer-to-Peer based k-anonymity location privacy service 
\cite{beresford2003location,gruteser2003anonymous,shokri2013hiding}, in which a user can borrow nearby other users' 
identity information to hide herself in a crowd. 
\textcolor{black}{A local mobile cloud (e.g., \cite{marinelli09hyrax,huerta2010virtual})
that relies on a cluster of nearby smartphones as computation resource providers can also be regarded as 
a type of DMACS system.}
Those services 
are provided by some smartphone users and consumed by some other smartphone users who are  
in close proximity of each other (e.g., within Bluetooth or Wi-Fi Direct
range of each other)
in a physical area (e.g., a coffee shop, a train station, or a sitting area in a public park). 
The data communication part of such services is supported by the ad-hoc networks
formed by smartphones. 
Furthermore, users dynamically arrive at and depart from the area over time.
We use the term \textit{DMACS system} to collectively denote such users, their smartphones,
the services, and the mechanism that incentivizes those users to participate in the provision and consumption
of those services. 
We expect that there will be substantial growth of DMACS applications as smart devices become
more powerful and pervasive in our daily lives. 

We emphasize that in a DMACS system, the services are provided by some users and used by some other users.
Since one cannot expect that users will always voluntarily contribute their limited resources of their smartphones 
(e.g., data plan quota, limited battery power and connection bandwidth) to support such services,
it is crucial to provide incentives (e.g., monetary value) for users who contribute their resources,
and in the mean time, those users who want to use such services should expect to pay a fair price for them.
A mechanism (that consists of a collection of algorithms, e.g., implemented in an app on smartphones) 
is needed to enable and facilitate the matching and transaction between users. 
For example, an incentive mechanism can help 
a user who is seeking a video streaming service find another user
who is willing to offer help (by providing a certain amount of data downloading capacity), and they  
can reach an agreement on a price/payment that is acceptable and fair to both parties.
Thus, it is essential to design an effective incentive mechanism to ensure the success of such a system,
which is the central research problem studied in this paper. 

\begin{figure}[htb!]
\centerline{
    \begin{minipage}{3in}
      \begin{center}
        \setlength{\epsfxsize}{1.6in}
\epsffile{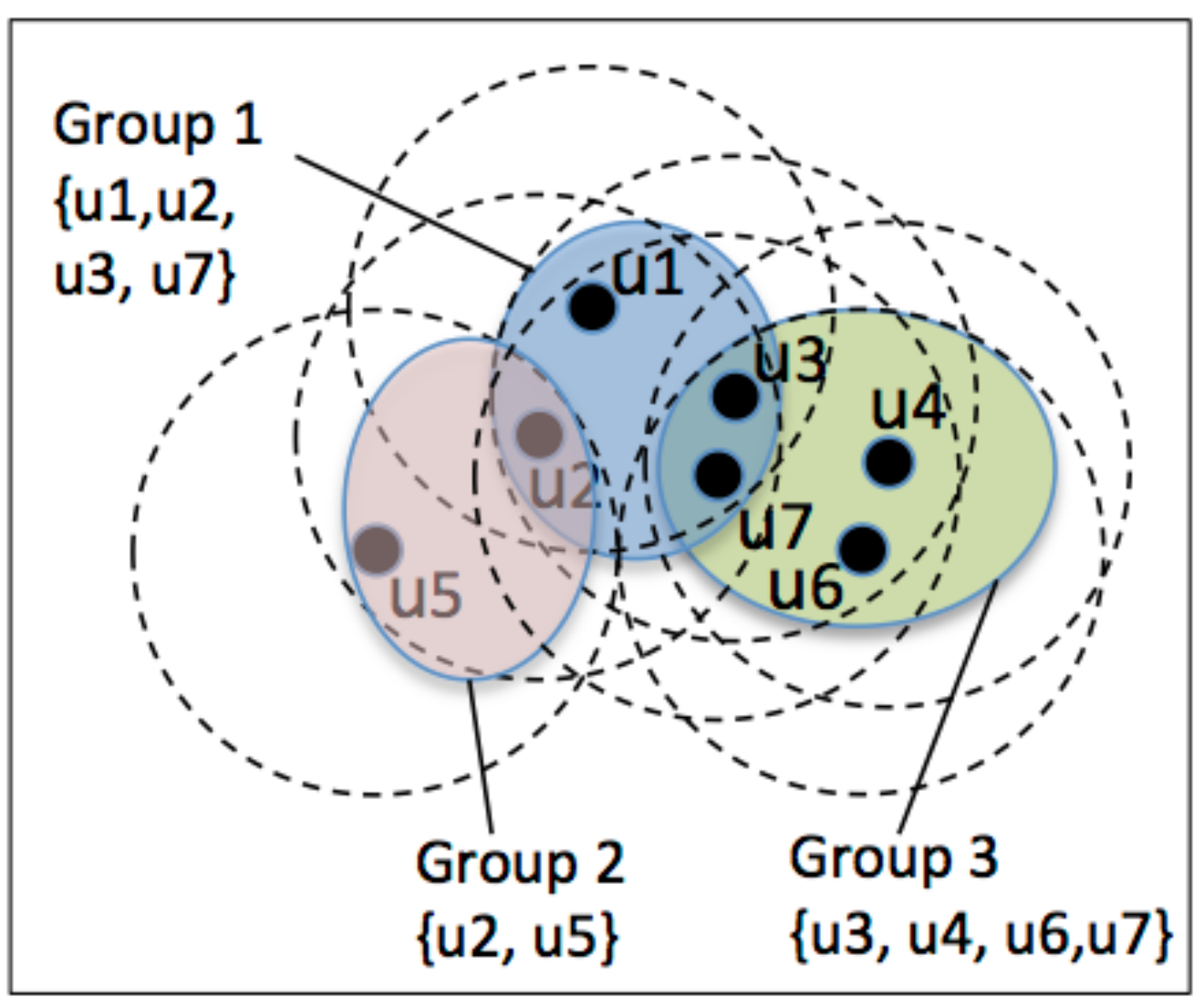}\\
       {}
      \end{center}
    \end{minipage}   
}
\caption{Multiple overlapping groups during a time period. 
Each dotted circle 
represents the radio range of a user centered at the circle. }
\label{fig.multigrp}
\end{figure}

In this paper, we design an auction-based incentive mechanism for a DMACS system. 
We first use the following example to describe the system, and based on 
which, we give a brief overview of our auction design.  
For example, while sitting in a park, a user might want to watch high-quality videos with the 
help from other users who have extra cellular data capacity to share. Or a user might 
want to use a k-anonymity location privacy service which relies on the 
help from other users around in the same park. In this scenario, the user 
is a service consumer or buyer and those other users who help her are service providers or sellers. 
In the mean time, if there are multiple other 
buyers also present in the park and try to purchase the service from those sellers, 
there will be a competition among multiple providers and consumers. 
In addition, users can arrive and depart over time, resulting in dynamic membership of the system,
leading to a challenge in system design. Note that the system 
can last a long time interval (e.g., a whole day).  
We can discretize this dynamic system by dividing 
the long time interval into multiple consecutive time periods (e.g., each being a half-hour),
and study the interaction among users in each period. 
Furthermore, those users co-existing in the same area during a single period
might not be within the wireless radio ranges of all other users. 
\textcolor{black}{That is, not every two users can directly communicate with each other, 
which is a common scenario due to the limitations of wireless signal strength.}
Think of a user as a node in a graph and the radio link between two users as an edge. 
We use \textit{group} to denote a complete graph in which every two users 
can communicate between themselves via their smartphones' Bluetooth or Wi-Fi Direct.
During each period, there might be multiple overlapping groups.
For example, in Figure \ref{fig.multigrp}, 
there are four groups: Group 1 consists of users $\{u_1, u_2, u_3, u_7\}$; Group 2 consists of 
users $\{u_2, u_5\}$; and Group 3 consists of users $\{u_3, u_4, u_6, u_7\}$.
A user can be in multiple groups at the same time. 
For example, in Figure \ref{fig.multigrp}, user $u_2$ is in Groups 1 and 2, as $u_2$ 
can communicate with $u_5, u_1, u_3, u_7$.

Note that a DMACS system cannot be described by the models in existing work on
\textit{dynamic double auction (DDA)} (e.g., \cite{bredin07chain,hajiaghayi2005online}), 
as they (to the best of our knowledge) always assume that users are in the same market or group. 
\textcolor{black}{
The nature of challenges faced in designing DMACS system calls for the design of a multi-market auction,
which is fundamentally  different from existing single-market auctions. 
In fact, a direct application of existing truthful single-market double auctions (e.g., McAfee auction \cite{mcafee1992dominant}) to each 
individual market (out of several overlapping markets)
cannot ensure users' truthfulness (which is a basic requirement for an auction). An example is described in detail in Section \ref{sec_mult_grp_example}}.
A pioneering work by Yang \textit{et al.} \cite{yang13location} applies double auction mechanism 
to assist k-anonymity location privacy service, but \cite{yang13location} only studies single period
and single-market auction mechanisms.  
In addition, a DMACS system is different from existing models on 
crowdsourcing to smartphones (e.g., \cite{xue_mobicomm,koutsopoulos2013optimal})
in that they consider the scenario where a server (working as an auctioneer) purchases
services (through the Internet) from smartphone users located in large geographical areas.

In this paper, we 
design a \textit{multi-market, dynamic double auction (M-DDA)}
as an incentive mechanism for a DMACS system.
We do not study the specific application services \textcolor{black}{(e.g., data sharing \cite{keller2012microcast}, 
location privacy protection \cite{shokri2013hiding}, or 
local mobile clouds \cite{marinelli09hyrax,huerta2010virtual,fernando2013mobile}})
that are supported by a DMACS system. Instead, we focus on the underlying
mechanisms that ensure the success of a DMACS system through incentivizing and facilitating users to collaboratively  
provide and benefit from those services. 
Our auction design consists of a collection of computationally efficient algorithms, which we implement
in a mobile app on Android platform\footnote{Our preliminary implementation is done
on Android Lollipop, which supports Bluetooth Low Energy (BLE) peripheral mode.
It can also be implemented on iOS platform which has iBeacon
capability. 
BLE can transmit to a range of up to 50 meters  
which is sufficient for the services we are interested in.}, referred to as \textit{MobiAuc}, 
working as an auctioneer.
\textcolor{black}{
Note that the Bluetooth Low Energy (BLE) on Android Lollipop allows the MobiAuc on a user's smartphone to broadcast 
messages (in peripheral mode) and scan for other phones' BLE advertisements even when 
the smartphone is in standby mode, with very low power consumption \cite{BLE-low-power,BLE-standby}. }

\begin{figure}[htb!]
\centerline{
    \begin{minipage}{3.3in}
      \begin{center}
        \setlength{\epsfxsize}{1.8in}
\epsffile{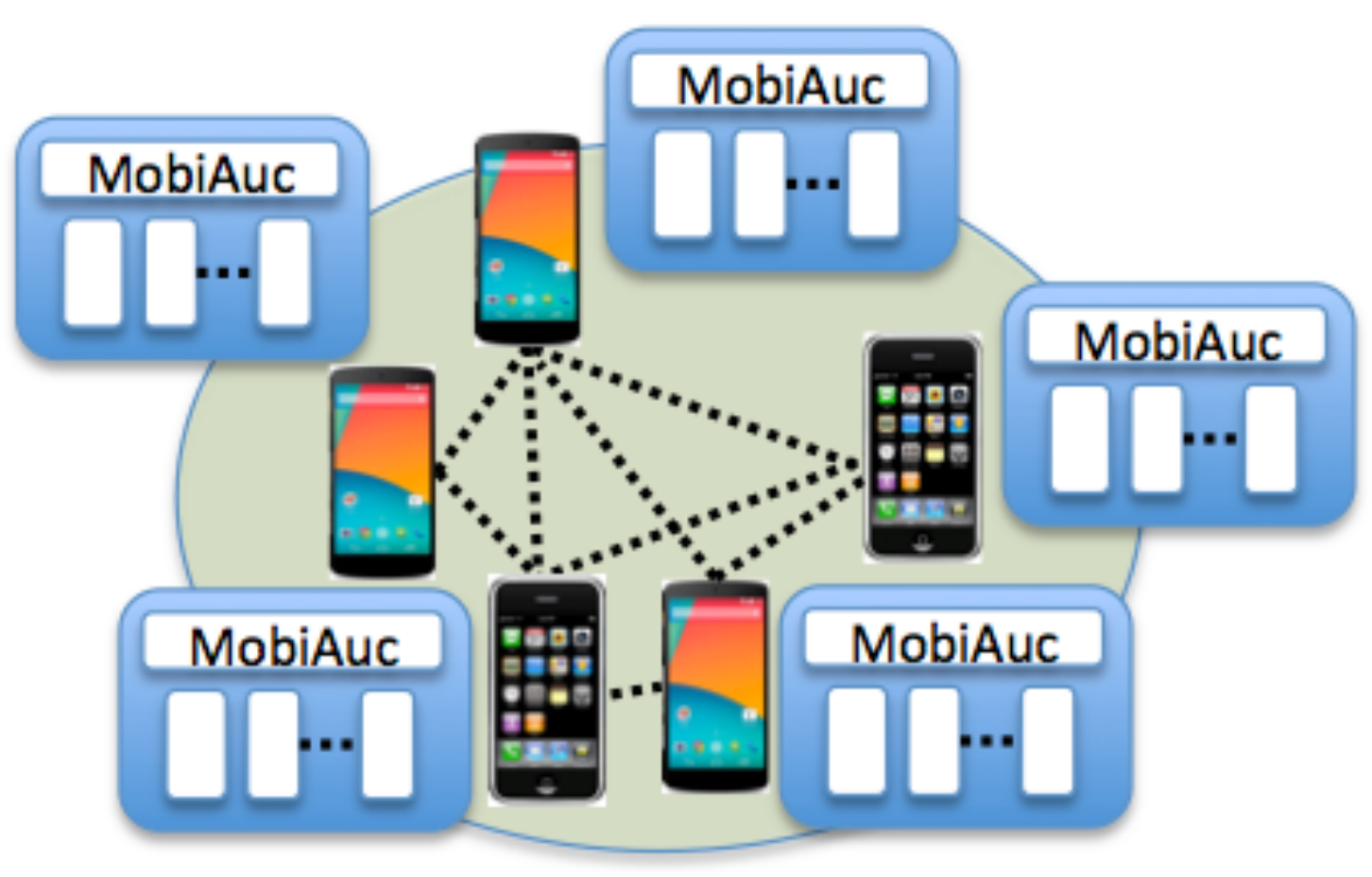}
        \setlength{\epsfxsize}{1.4in}
\epsffile{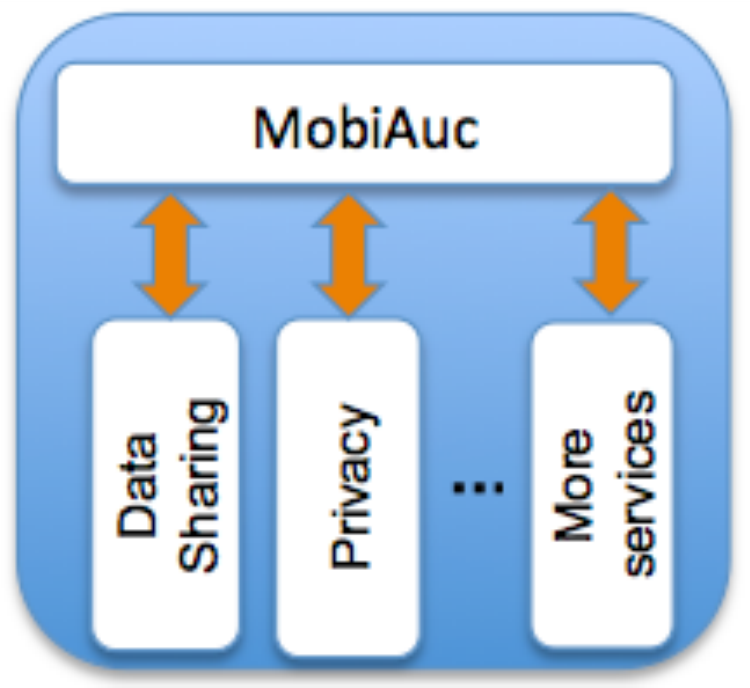}\\
      \end{center}
    \end{minipage}  
}
   \caption{
   \textcolor{black}{Example scenario of M-DDA for DMACS. MobiAuc apps (implementing M-DDA) on smartphones
   select users for trade, \textit{i.e.}, the provision/consumption of DMACS services such as data sharing, location privacy, etc. 
   M-DDA takes place only in the ad-hoc network formed by users. In addition, 
   MobiAuc apps  also communicate with backend servers for other
   system services such as auditing.}
   }   
  \label{fig.mobicrowd_archit}
\end{figure}

We now utilize the scenario shown in Figure \ref{fig.mobicrowd_archit}
to put our incentive mechanism in a practical context. 
As shown in the figure,
the MobiAuc apps installed on those smartphones (carried by users) 
compute bids/asks, and communicate those bids/asks with 
each other through beaconing via \textcolor{black}{BLE, a process similar to the flooding  
of link state information in a link-state routing protocol}.
Those apps also communicate with backend servers in cloud for other system services such as  
transaction auditing, etc.
The auction algorithms run by the MobiAuc apps play a central role in this system. 
This paper focuses on the design of those algorithms.

We make the following important contributions.

\begin{itemize}

\item 
We design an auction-based incentive mechanism for the emerging user-assisted mobile crowd services, 
enabled by the latest smartphone technology. 
This mechanism includes an effective auction design with key components such as
a well-defined single-period matching rule \textcolor{black}{(referred to as Virtual Market matching rule)}, 
and \textcolor{black}{group matching} algorithm, etc.
Our auction design satisfies desired properties of an effective dynamic auction, \textit{i.e.}, it is 
truthful (\textit{i.e.}, no user can benefit from cheating about her true valuation of the service,
arrival time, group membership, etc.), feasible, individual-rational, 
no-deficit, and computationally efficient. 

\item A main novelty of our auction design is that it 
addresses the \textcolor{black}{overlapping} multi-group or multi-market nature of a DMACS system
(\textit{i.e.}, multiple overlapping groups co-existing in a time period). 
The multi-group characteristic (inherent in mobile wireless environment) 
distinguishes our auction from existing dynamic double auction designs  
(which assume that users are in the same auction market).
\textcolor{black}{
Note that a direct application of existing truthful single-market double auctions (e.g., McAfee auction \cite{mcafee1992dominant}) to each 
individual market (out of several overlapping markets)
cannot ensure users' truthfulness.}

\item Our extensive simulations based on generated user patterns
(stochastic arrivals, random grouping and clustering of users)
and real-world traces 
demonstrate that our auction algorithms achieve good system efficiency. 
We also show that even in an environment of multiple overlapping groups, 
our design can achieve an efficiency level  
that is comparable to that of a truthful auction in 
a single-group environment,
if each user can be in a moderate number of randomly chosen groups simultaneously. 

\item Furthermore, our evaluation shows that the \textit{price of truthfulness guarantee}
(\textit{i.e.}, efficiency loss of a truthful auction design when compared against a
non-truthful online random greedy algorithm)
can be large. 
This result is consistent with the impossibility result by Myerson and Satterthwaite \cite{myerson1983efficient},
and our results further indicate a high price we need to pay in order to have guaranteed truthfulness
in a multi-market dynamic double auction.  

\end{itemize}

The rest of the paper is organizes as follows.
In Section \ref{sec_sys}, we present our system model. 
In Section \ref{sec_auction} we describe our proposed auction mechanism.
We evaluate our auction design in Section \ref{sec_evaluation}. Related work is given 
in Section \ref{sec_related}, and the paper concludes in Section \ref{sec:conclusion}.

\section{System Model}\label{sec_sys}

\subsection{Preliminaries}

A dynamic, mobile ad-hoc crowd service (DMACS) system can be described as 
a dynamic system discretized in a sequence of time periods $T={1, 2, 3, ...}$, indexed by $t$. 
Let $a_i$ denote user $u_i$'s arrival time, and $d_i$ the departure time.
We assume a user's stay duration in the system ($d_i-a_i$, referred to as \textit{patience}) is bounded by a constant $K$
(referred to as \textit{maximum patience}),
that is, $d_i\le a_i + K, \forall i$. 
Consider those users present in the system in a particular period $t$.
We can think of a user as a node in an undirected graph at $t$. If two users are within each other's
wireless radio range, then there exists an edge between them in the graph. 
The graph of those users can possibly be partitioned into several connected components. 
A connected component can be a complete graph by itself, referred to as a \textit{group}.
Or it might not be a complete graph by itself (\textit{i.e.},
there exist at least two users who do not have a direct edge between themselves),
and such a connected component is
essentially a cluster 
of multiple overlapping groups (or maximal cliques). Figure \ref{fig.multigrp} shows such an example.

If a user is honest about which group(s) she is in, 
then her group membership information is only a result of her device's radio transmission range. 
However, 
a user might want to lie about her group membership in order to gain advantage in 
service trading (if there is no mechanism in place to ensure truthfulness from users). 
A user's cheating behavior will be discussed later. 
Let vector 
$\mathbb{G}(t)=\langle g^1(t), g^2(t), ..., g^\ell(t) \rangle$
denote the set of all groups existing at time $t$ (where $\ell$ is the number of groups at $t$). 
Let binary vector
$G_i(t)=\langle g^1_i(t), g^2_i(t), ..., g^\ell_i(t) \rangle$
indicate the groups where user $u_i$ is in at time $t$. That is, $g^j_i(t)=1$ means user $u_i$
is in group $j$ at time $t$; $g^j_i(t)=0$ otherwise. 
Let $\mathbf{G_i}(a_i, d_i)$ denote the vector of groups where user $u_i$ is in 
during her life time in the system, \textit{i.e.}, 
$\mathbf{G_i}(a_i, d_i)= \langle G_i(a_i), G_i(a_i+1),..., G_i(d_i) \rangle$.

Users who purchase a mobile service are referred to as \textit{buyers}, 
whereas users who offer the mobile service are referred to as \textit{sellers}.
\textcolor{black}{
For ease of presentation, we introduce our model for 
homogeneous service environment where the levels of the qualities of services
offered by all sellers are the same.
Our model and algorithms are also applicable to heterogeneous service environment,
as discussed in Section \ref{sec_hetg}.}
A buyer can do a transaction (\textit{i.e.}, receiving service from) with only one seller in a period $t$, 
and vice versa. 
When a pair of buyer-seller is chosen for trading, 
the amount of service traded between the pair of buyer-seller is identical to that of any other buyer-seller pair.
The payment offered by a buyer is referred to as \emph{bid}, and the price 
asked by a seller is referred to as \emph{ask}.
Let $\mathbb{B}(t)$ and $\mathbb{S}(t)$ denote respectively the set of buyers 
and the set of sellers who \textit{arrive} in period $t$. 
Let $\mathcal{B}(t)$ and $\mathcal{S}(t)$ denote respectively the set of \textit{active} 
buyers and sellers in period $t$ (note that a user is regarded as active if she satisfies some requirements,
which will be discussed later).
Let $B(t)\in \mathbb{R}^m_{> 0}$ and $S(t)\in \mathbb{R}^n_{\ge 0}$ 
denote the set of active bids and active asks in period $t$,
where $m\ge 0$ and $n\ge 0$ denotes respectively the number of bids and asks.
\textcolor{black}{
We use $u_i$ to represent a user (who can be a buyer or a seller).  
We use $u^b_i$ to explicitly represent a buyer when necessary, and $b_i$ denotes her bid.  
Similarly, $u^s_j$ represents a seller when necessary, and $s_j$ represents her ask. 
Note that $a_i =t, \forall u^b_i \in \mathbb{B}(t)$, and  $a_j =t, \forall u^s_j \in \mathbb{S}(t)$. 
Let $w_i \in [0,+\infty) $ denote a user's true valuation 
of the mobile service; $w_i=b_i$ for buyer $u^b_i$; $w_j=s_j$ for seller $u^s_j$.
A user's true \textit{type} is specified by 
 \begin{equation}
 \theta_i=(\mathbf{G_i}(a_i, d_i), w_i)
 \end{equation}
Buyer $u^b_i$'s true type is $\theta_i=(\mathbf{G_i}(a_i, d_i), b_i)$.
Seller $u^s_j$'s true type is $\theta_j=(\mathbf{G_j}(a_j, d_j), s_j)$.}

\subsubsection{System requirements}

We model a DMACS system as a type of double auction \cite{mcafee1992dominant}, and 
assume that there is an auctioneer that determines which buyers and sellers will 
be selected for trade or service transaction.
\textcolor{black}{We want to emphasize that the auctioneer in our system 
is not a real person or a central entity physically located in some place in our system. 
Instead, the auctioneer refers to a collection of identical apps, e.g., the MobiAuc, that run our auction algorithms and reside in users' smartphones.
Since all those auctioneer apps work on the same information of the whole system and they use the same auction algorithms, 
they will produce the same auction result. This is similar to a link-state routing protocol (e.g., OSPF) that resides in each router
in a network
and runs the same algorithm on the same link-state information of the network.}
We now explain the system setting from an auction perspective,
and leave a formal description of our auction design in later sections.  

The auctioneer app in each phone periodically broadcasts 
beacons (e.g., via BLE peripheral mode \cite{ble})
that contains the phone owner's type information, 
and receives beacons from other phones.
A user is given the option to enter in her app her information such as 
ask or bid, or she can delay beacon broadcast to change her 
arrival time in the system.

Once a user joins the system, 
the user's app broadcasts her value, arrival and departure time 
to other users.  
Her app also broadcasts the list of other users that it can receive. 
Once her app receives other users' broadcast lists, her app knows which group(s) she 
is in (unified naming of group IDs can be easily done with a few more negotiation beacons). 
Note that during a single-period auction (to be discussed later),
before the auctioneer has determined who will be the winning buyers and sellers,
a user can broadcast her information repeatedly, but her information 
in those messages should be consistent. For example, the user cannot say her 
ask is $\$2$ in one message and change it to  $\$3$ in later message(s). 
This restriction is necessary to prevent a user from changing her type based 
on the received types of other users.
\textcolor{black}{The consistency of a user's type information in her messages 
can be easily verified by examining whether the secure hash (e.g., SHA-2)
of her type information (attached to each beacon message) is consistent across her messages.}
Besides broadcasting its own information,
an auctioneer app also re-broadcasts periodically the information it has received from other users.
Note that a message cannot be falsified or altered, which 
can be easily guaranteed by secure hash and public-key encryption for strong identity purpose. 
This beacon relaying is necessary to allow all apps in a cluster to have an unified 
view of the cluster. 
\textcolor{black}{The group membership information of a user during her stay in the system 
can be easily determined and updated by her auctioneer app by reading received messages from other smartphones.
A user does not need to get involved in this group membership determination process, unless the user
wants to cheat about her group membership.}
 
\subsubsection{Cheating, reported type, and utility}\label{sec_cru}

We assume that a user may want to cheat about her type to gain benefit.
For example, a buyer might want to report a lower bid than the amount that she actually can pay, 
in order to pay less for the same service. 
Below is a summary of user cheating behaviors.

\begin{enumerate}
\item A user can report an arrival time later than its true arrival time
(by delaying her broadcast beacons); but she cannot report an earlier arrival time
(due to radio range constraint). This behavior is also considered in
\cite{bredin07chain,hajiaghayi2005online}. 
 
\item A user $u_i$ can report an earlier or later departure time than $d_i$,
once she arrives in the system. 

\item A user  $u_i$ can report a lower or higher value than $w_i$.

\item A user $u_i$ can report that she is 
in a \textit{fewer} number of groups than she is actually in, 
but she cannot report that she is in more groups. 
For example, in Figure \ref{fig.multigrp},
$u_3$ can broadcast that she can only receive beacons from 
$u_4, u_6$ and $u_7$, which means $u_3$ reports that she is only in Group 3,
even though she is actually also in Group 1.
Even though $u_1$ and $u_2$ can hear $u_3$, but they choose to believe $u_3$'s
report as it is possible that $u_3$ simply just has a bad radio receiver. 
However, a user cannot report that she is in more groups, as it 
can be easily detected by other users through cross-checking their 
received reports (assuming there is no collusion). In addition, once a user is 
selected as a winner, she
should fulfill her promise such as giving a payment to the auctioneer or 
providing service to a buyer, thus,
reporting a group where she is not actually in can be easily detected. 

\end{enumerate}

Let $\hat{a}_i$ and $\hat{d}_i$ denote respectively the reported arrival time
and departure time of user $u_i$. 
Let $\hat{b}_i$
denote the reported bid of buyer $u^b_i$; and $\hat{s}_j$  the reported ask
of seller $u^s_j$.
Let $\mathbf{\hat{G}_i}(\hat{a}_i, \hat{d}_i)$ denote $u_i$'s reported group. 
Then, $u_i$'s reported type is denoted by 
\begin{equation}
\hat{\theta}_i=(\mathbf{\hat{G}_i}(\hat{a}_i,\hat{d}_i), \hat{w}_i),
\end{equation}
where $\hat{w}_i$ can be  $\hat{b}_i$ (if $u_i$ is a buyer) or $\hat{s}_i$ (if $u_i$ is a seller). 

The auctioneer 
chooses a set of winning buyers and a set of winning sellers,
and decides the payment that those winning buyers should pay to the auctioneer
and the payment that those winning sellers will receive from the auctioneer. 
Let $p_i$ denote the payment for user $u_i$. 
As in \cite{bredin07chain,yang13location,hajiaghayi2005online}, we model risk-neutral agents with quasi-linear
utility. Buyer $u^b_i$'s utility is given by 
$U(u^b_i) = b_i - p_i$ if she is selected as a winner; 
$U(u^b_i) = 0$ otherwise. 
Seller $u^s_j$'s utility is given by 
$U(u^s_j) = p_j - s_j $ if she is selected as a winner; 
$U(u^s_j) = 0$ otherwise.

\subsection{Multi-market dynamic double auction}

We model the incentive mechanism of a DMACS system as a dynamic double auction 
for a collection of multiple overlapping markets or groups 
(e.g., there are $3$ groups in Figure \ref{fig.multigrp}). 
We call such a model Multi-market DDA or M-DDA. 
To the best of our knowledge, existing dynamic double auctions (e.g., \cite{bredin07chain,hajiaghayi2005online}) 
only consider a single market (\textit{i.e.}, all buyers and sellers 
that are present in a time period join the same auction market).
An M-DDA consists of a sequence correlated auctions, each taking place in 
a single period.
A single-period or single-round auction is a direct-revelation double auction (e.g., \cite{mcafee1992dominant,yang13location})
in which a user reports to the auctioneer
only a single direct claim about its type $\hat{\theta}_i$.
The auction is dynamic as users join or leave the system over time. 

In this paper, we introduce a design of M-DDA, referred to as   
M-CHAIN auction.
M-CHAIN is built on top of the single-market CHAIN framework 
proposed in \cite{bredin07chain}.
\textcolor{black}{ 
Note that due to the nature of multi-group setup in our system, 
the algorithms discussed in \cite{bredin07chain} may not be applicable to our system. 
At the heart of our M-CHAIN mechanism is a well-defined single-period matching rule, referred to as 
Virtual Market (VM) matching rule,
which utilizes McAfee single-round auction \cite{mcafee1992dominant} as a subroutine.}
A problem instance that is solved by M-CHAIN can be formalized 
as a tuple  
\begin{equation}
\langle \mathbb{B}(T), \mathbb{S}(T), \mathbb{G}(T), T, K\rangle.
\label{eqn_prob_inst}
\end{equation}
Note that $T$ is a sequence of periods.
With a slight abuse of notation, we also let $T$ represent the maximum period of the system.
M-CHAIN differs from the ones in \cite{yang13location} in that \cite{yang13location}
does not deal with a multi-market situation, nor a dynamic system. 
\textcolor{black}{Next we give a brief review of CHAIN, McAfee Auction \cite{mcafee1992dominant},
and some other techniques utilized in our M-CHAIN design.}

\subsubsection{A brief review of CHAIN framework}

CHAIN \cite{bredin07chain} is an algorithmic framework that can be used to construct truthful dynamic double 
auctions through the plug-ins of various modules,
with an underlying assumption that all users are in a common market.
Its basic algorithmic flow is as follows.
During each period $t$, the auctioneer first decides whether to accept a newly arrived user $u_i$
by generating an admission price $q_i(t)$ for the user. 
If $u_i$ is a buyer and $\hat{b}_i \ge q_i(t)$, then $u_i$ is admitted; if $u_i$ is a seller
and $\hat{s}_i \le q_i(t)$, then $u_i$ is admitted.
Then the auctioneer form a set of \textit{active} buyers 
and a set of \textit{active} sellers for time $t$, which consist of newly accepted users and those
who were not chosen by the auctioneer as winners in period $t-1$ but are allowed
by the auctioneer to stay in the system till time $t$ (as they satisfy some requirements, discussed below). 
Then, the auctioneer runs a single-period matching rule to choose
a set of buyers and a set of sellers as winners in period $t$, out of all \textit{active}
buyers or sellers in period $t$. 
The auctioneer also constructs a Strong-No-Trade (SNT) set of users for period $t$ (denoted by $SNT(t)$), and 
only those losing user who are in $SNT(t)$ will be survivors (\textit{i.e.}, \textit{active} at $t+1$). 
After a single-period auction at time $t$, a user can be in one of the three states:
winning, survived (\textit{i.e.}, in $SNT(t)$), or priced-out.  

The three key components of CHAIN are: a well-defined single-period matching rule, 
a valid SNT set construction for each period $t$, and a procedure to calculate the admission price $q_i(t)$.
A single-period matching rule is an auction algorithm that 
selects winning buyers and sellers out of active buyers and sellers, and 
determines the payment that those winning buyers should pay to the auctioneer
and the payment that those winning sellers will receive from the auctioneer.
A SNT set construction for a matching rule in time $t$ 
is valid if the users in the set satisfies the following three requirements (we take user $u_i$ as an example): 
(1) user $u_i$ won't be chosen for trade no matter what value $\hat{w}_i$ she reports
as long as $\hat{w}_i\in \mathbb{R}$ and other users' reported values remain unchanged; (2) 
while $\hat{d}_i>t$,  user $u_i$ cannot manipulate $\hat{\theta}_i$ to affect whether herself
is chosen into the SNT set, nor can she affect whether other users will be put into the SNT set.
For a formal definition of a valid SNT, see \cite{bredin07chain}.  
Another requirement of CHAIN is that all users's \textit{patiences}, 
should be bounded by a constant $K$ (the maximum patience).
A formal definition of a single-period well-defined matching rule is given in \cite{bredin07chain} 
and re-stated below.
\begin{theorem}\cite{bredin07chain}
\textbf{Single-period well-defined matching rule}: 
A single-period matching rule is well-defined if it satisfies the following properties:
truthfulness, no-deficit, individual-rationality, and feasibility.
\label{def_mr}
\end{theorem}

The above four properties are explained below:
\begin{itemize}
\item \textit{Truthfulness}, no user can benefit from cheating about her true type $\theta_i$ 
(\textit{i.e.}, $\hat{\theta}_i\ne \theta_i$).
\item  \textit{Individual rationality}, a buyer or seller 
always receives a nonnegative utility by participating in the auction and reporting her true type.
\item  \textit{No-deficit or budget balance},
the total payment collected from the winning buyers is at least as large as the total payment paid to the winning sellers.
\item  \textit{Feasibility}, each user's service transaction can take place 
within her reported life time in the system if she is selected as a winning buyer or seller. 
\end{itemize}

Note that we differentiate an auction's \textit{system efficiency} from 
its \textit{computation efficiency}. 
\textit{System efficiency} represents the 
quality of an auction's outcome when it is compared against the outcome 
of an optimal offline algorithm assuming all users are truthful.
Its formal definition is given in Section \ref{sec_evaluation}.
The impossibility result by Myerson and Satterthwaite \cite{myerson1983efficient} 
says that it is impossible to simultaneously achieve maximum system efficiency, truthfulness, 
individual rationality, and no-deficit.
\textit{Computation efficiency} represents the time complexity of an auction algorithm.

Bredin \textit{et al.} \cite{bredin07chain} have shown that an online or dynamic double auction (DDA) algorithm
constructed from CHAIN framework has all desired properties of an auction.
\begin{lemma}  \cite{bredin07chain}
A dynamic double auction (DDA) CHAIN algorithm is truthful, no-deficit, individual-rational, and feasible,
when it uses a well-defined single-period matching rule, a valid SNT construction, 
and users' patiences are bounded by a constant $K$. 
\label{lemma_da_chain}
\end{lemma}

A problem instance for a CHAIN-based auction algorithm can be described as
$\langle \mathbb{B}(T), \mathbb{S}(T), g_0(T), T, K \rangle$,
where $g_0(T)$ represents 
the only group that exists in each period. 
\textcolor{black}{Note that CHAIN's problem instance construction is a special case
of that for our proposed M-CHAIN, as $\mathbb{G}(T)$ in M-CHAIN's problem instance (\ref{eqn_prob_inst}) indicates
that in M-CHAIN, there can exist multiple groups in each period.}

In addition, our auction design uses McAfee auction \cite{mcafee1992dominant} as a subroutine.
We briefly review this auction as follows. Since McAfee auction algorithm deals with single period only,  
we ignore the time notation $t$ here. 

\subsubsection{McAfee auction algorithm \cite{mcafee1992dominant}} \label{sec_mcafee}
If $\min(|B|,|S|)<2$, then there is no trade. 
Otherwise, place two dummy bids (and asks) with values $\infty$ and $0$
into $B$ and $S$. Sort those bids and asks in descending and ascending order, respectively,
with ties broken arbitrarily.  
Let $b_{(0)} \ge b_{(1)} \ge ... \ge b_{(m)}$ and $s_{(0)} \le s_{(1)}  \le ...  \le s_{(n)}$ after the sorting.
Note that $(b_{(0)}, s_{(0)})$ denotes dummy pair $(\infty, 0)$ and 
$(b_{(m)}, s_{(n)})$ denotes dummy pair $(0, \infty)$. 
Let $k \ge 0$ index the last pair of bids and asks
such that $b_{(k)} \ge s_{(k)}$ (hence $b_{(k+1)} < s_{(k+1)}$). 
Let $p_{(k+1)} = (b_{(k+1)}+ s_{(k+1)})/2 $.
When $k \ge 1$, consider the following two cases:
\textbf{Case I}. If price $b_{(k)} \ge p_{(k+1)}$ and $s_{(k)} \le p_{(k+1)}$ then the first $k$ bids and 
asks trade and payment $p_{(k+1)}$ is collected from each winning buyer and made to each winning seller.
\textbf{Case II}. Otherwise, the first $k-1$ bids and asks trade; $b_{(k)}$ 
is collected from each winning buyer; $s_{(k)}$ is made to each winning seller.

\subsubsection{Single-market truthful auction cannot ensure truthfulness in multi-market scenarios}\label{sec_mult_grp_example}

\begin{figure}[htb!]
\centerline{
    \begin{minipage}{1.5in}
      \begin{center}
        \setlength{\epsfxsize}{1.4in}
\epsffile{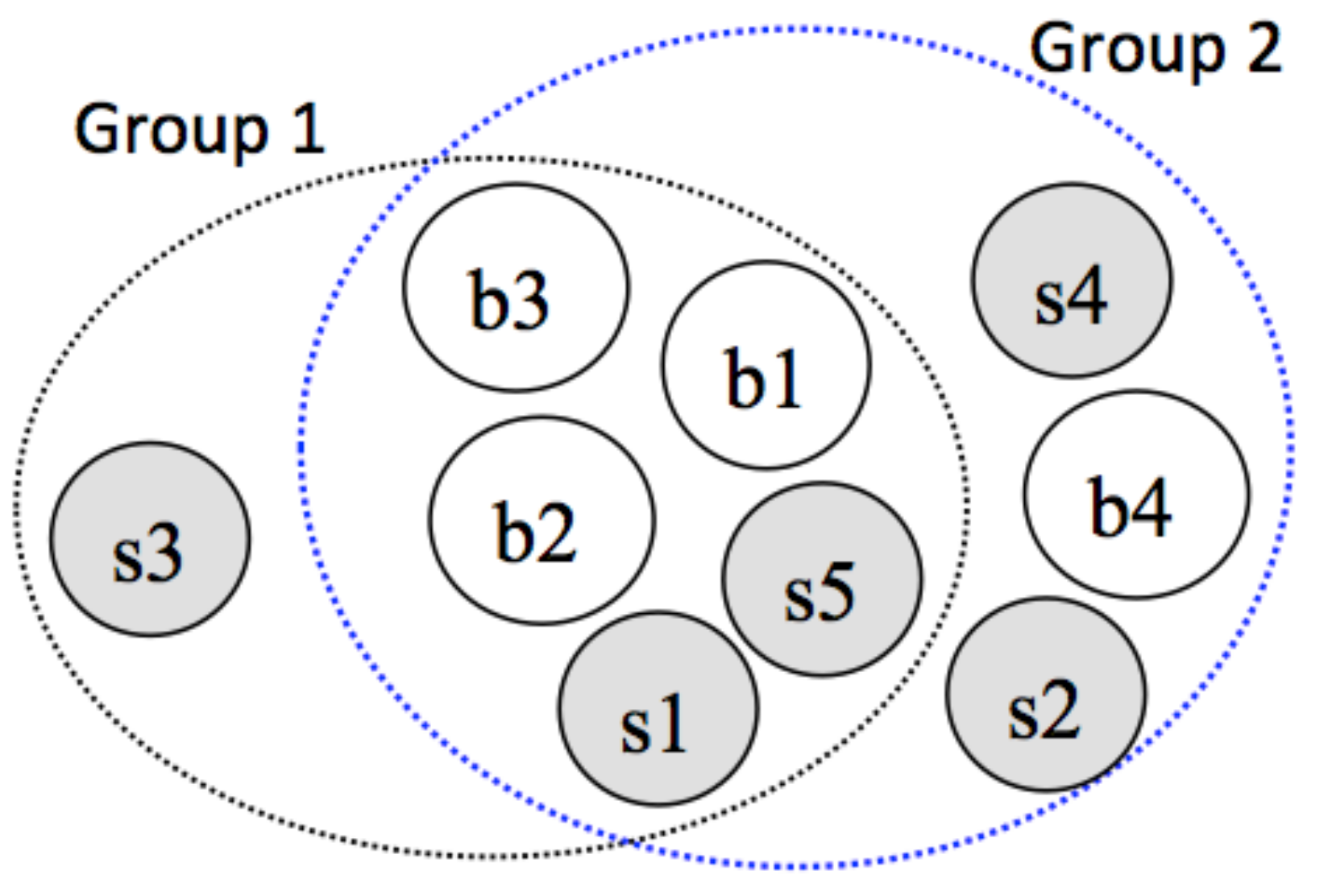}\\
       {}
      \end{center}
\caption{The true group membership of those users (represented by their bids/asks).}
\label{fig.multi_grp_1}
    \end{minipage}  
    \hspace{4pt} 
    \begin{minipage}{1.5in}
      \begin{center}
        \setlength{\epsfxsize}{1.4in}
\epsffile{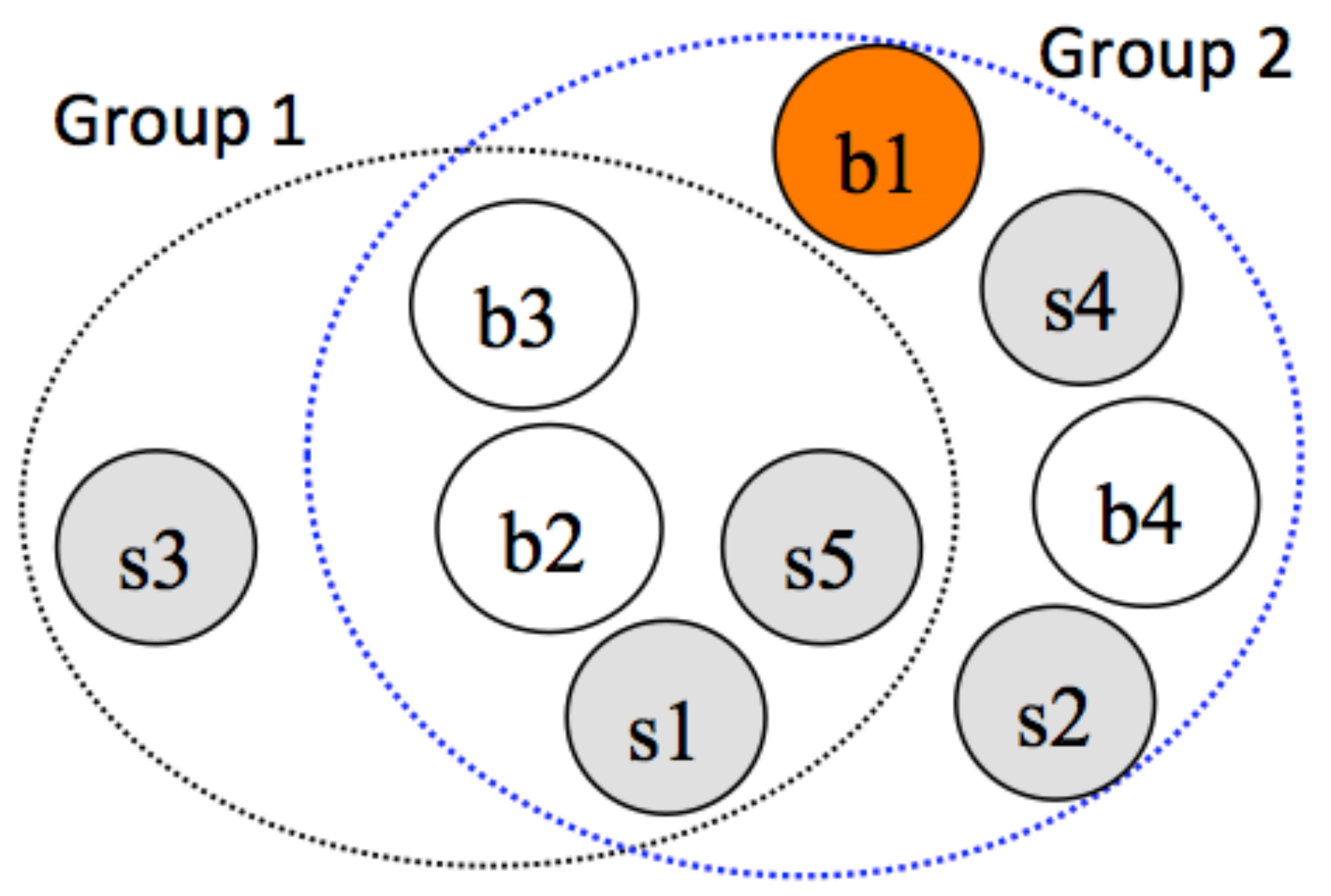}\\
       {}
      \end{center}
\caption{The broadcast message sent by buyer $u_1^b$
does not say that she can directly communicate with seller $u_3^s$.}
\label{fig.multi_grp_2}
    \end{minipage}  
    \hspace{4pt} 
}
\end{figure}

\textcolor{black}{
We now use an example to illustrate that directly applying McAfee auction (a single-market auction) 
in each individual group or market cannot ensure truthfulness from users in a multi-market scenario.
Consider four buyers $\{u_1^b, u_2^b, u_3^b, u_4^b\}$ and five sellers $\{u_1^s, u_2^s, u_3^s, u_4^s, u_5^s\}$. 
their true bids and true asks are
$\{b_1 = \$12, b_2= \$10, b_3=\$2, b_4=\$1 \}$ and $\{ s_1=\$1, s_2=\$1, s_3=\$4, s_4=\$3, s_5=\$5 \}$.
There are two groups in the system: Group 1 and Group 2. A  user belongs to either one of the two groups, or belongs to both groups. 
The \textit{true} group membership of those users is illustrated in Figure \ref{fig.multi_grp_1}.
Suppose that the auction in Group 1 is conducted first, and it is followed by the auction in Group 2. 
Both groups conduct McAfee auction.
If all users report their true group membership and true bids/asks, then $b_1$ and $s_1$ will be selected
as winners during Group 1's auction ($u_1^b$'s payment is $\$10$ and $u_1^s$ receives $\$4$) , 
and $b_2$ and $s_2$ will be selected as winners during Group 2's auction ($u_2^b$'s payment is $\$2.5$ and $u_2^s$ receives $\$2.5$).
Now consider that buyer $u_1^b$ cheats by saying that she is not in Group 1 (i.e., she is not aware of the existence of 
user $u_3^s$), illustrated in Figure \ref{fig.multi_grp_2}. Then, during Group 1's auction (conducted first), $u_2^b$ and $u_1^s$ are winners. 
During Group 2's auction, $u_1^b$ and $u_2^s$ are winners, and $u_1^b$'s payment is $\$2.5$, which  
is less than $\$10$ that she has to pay (when she does not cheat).
}

\section{Truthful Auction Design for DMACS System} \label{sec_auction}

We present in this section multi-market CHAIN, a truthful 
and computationally efficient auction design, as an incentive mechanism for a DMACS system.

\subsection{M-CHAIN auction}

Multi-market CHAIN (M-CHAIN) is displayed in Algorithm \ref{alg.mchain}.
At the heart of this auction are a well-defined single-period matching-rule, 
referred to as 
Virtual Market (VM) 
matching rule (shown as Algorithm \ref{alg.GPM}),
a valid strong no-trade construction for each period, 
and a truthful Group Matching algorithm (shown as Algorithm \ref{alg.gm}).
The strong no-trade set $SNT(t)$ 
consists of two sets $SNT_B(t)$ and $SNT_S(t)$, for buyers and sellers respectively.   
We use $\mathbb{H}(t) \in \mathbb{R}^h$ (\textit{i.e.}, the history in period $t$) to denote the set of users 
that were either priced out or selected (as winners) by the auctioneer at time $t$. Here $h$ is the total
number of users in $\mathbb{H}(t)$.

\begin{algorithm}
\caption{ \textbf{M-CHAIN}}
\label{alg.mchain}
\begin{algorithmic}[1]
\REQUIRE $\langle \mathbb{B}(T), \mathbb{S}(T), \mathbb{G}(T), T, K\rangle$
\STATE $t \leftarrow 0$

\WHILE{$t \le T$} 
	\STATE Active buyer set $\mathcal{B}(t) \leftarrow SNT_B(t-1)$ \\
		Active seller set $\mathcal{S}(t) \leftarrow SNT_S(t-1) $. 
	
	\FOR{ each user $u_i$ in $\mathbb{B}(t)$ and $\mathbb{S}(t)$} 
		
		\STATE {Calculate the admission price $q_i$ according to (\ref{eqn.admission}).}
		\STATE If $u_i$ is a buyer and $\hat{b}_i\ge q_i$, then add her into $\mathcal{B}(t)$.  
		 If $u_i$ is a seller and $\hat{s}_i\le q_i$, then add her into $\mathcal{S}(t)$.
		
	\ENDFOR
		\STATE Invoke the single-period matching rule given by Algorithm \ref{alg.GPM} 
		to select winning buyers and sellers from  $\mathcal{B}(t)$ and $\mathcal{S}(t)$,
		and determine their payments.
		\STATE Losing users that are not in $SNT_B(t)$ or $SNT_S(t)$ will be priced out and put into history $\mathbb{H}(t)$. 
\ENDWHILE
\end{algorithmic}
\end{algorithm}

The admission price calculation (given in \cite{bredin07chain}) is as follows. 
If $u_i$ would lose in $t'$ and $u_i\notin SNT(t'), \forall t' \in [\hat{d}_i-K, \hat{a}_i-1]$, then $q_i=\infty$ (for buyer)
or $q_i= - \infty$ (for seller);
otherwise, calculate  $q_i$ as follows:
\begin{equation}
q_i =
  \begin{cases}
   \max_{t' \in [\hat{d}_i-K, \hat{a}_i-1], i\notin SNT(t')} [p_i(t'), -\infty], & \mbox{for buyer} \\
   \min_{t' \in [\hat{d}_i-K, \hat{a}_i-1], i\notin SNT(t')} [p_i(t'), \infty],  & \mbox{for seller} 
  \end{cases}
  \label{eqn.admission}
\end{equation}
where $p_i(t')$ is the payment that a buyer would pay (or a seller would receive) in period $t'$
if she would join the auction with other users in $\mathbb{H}(t')$.
Such a payment is determined by the matching rule given by Algorithm \ref{alg.GPM}. 
A user $u_i$ is admitted if her patience equals $K$,
or $u_i \in SNT(t'), \forall t' \in [\hat{d}_i-K, \hat{a}_i-1]$.

\subsubsection{Virtual market single-period matching rule}

\textcolor{black}{The idea of 
Virtual Market (VM) matching
rule is as follows.
The auctioneer app on each user's smartphone conducts auction by pooling all sellers and buyers together into a 
single virtual market, without considering their group information. Since all auctioneer apps have the same knowledge 
after the message exchange phase of each period and they run the same auction algorithms, 
so they will derive the same auction result.
For example, MobiAuc apps running on users' smartphones obtain the same knowledge of the system through 
message exchanges via broadcast beaconing and scanning, a process similar to OSPF's link-state information flooding. 
Note that the MobiAuc app can continuously broadcast and scan via Bluetooth Low Energy (with very low power consumption), 
even when a device (e.g., Nexus 6) is in standby mode.
}
\textcolor{black}{The auctioneer apps run McAfee auction  \cite{mcafee1992dominant} (described in Section \ref{sec_mcafee}) 
for the virtual market
to get a candidate set of winners as candidates, denoted by sets $\mathcal{B}^c(t)$ (buyers) and $\mathcal{S}^c(t)$ (sellers),
and then from them, they use Group Matching (Algorithm \ref{alg.gm}) 
to select the sets of final winning buyers and sellers, denoted by $\mathcal{B}^W(t)$ and $\mathcal{S}^W(t)$.}

This rule is shown in Algorithm \ref{alg.GPM}. 
Since we focus on a single period here, we drop time notation $t$ for ease of exposition. 
Recall that $\mathcal{S}$ and $\mathcal{B}$ denote active seller set and active buyer set respectively.
Assume that there are $\ell$ groups at $t$. 
Let $G$ denote the set of all users from all groups.

\begin{algorithm}
\caption{ \textbf{Virtual Market} single-period matching rule} 
\label{alg.GPM}
\begin{algorithmic}[1]

\REQUIRE $ \mathcal{B}, \mathcal{S}, \mathbb{G}$. \\

\STATE $G \leftarrow \bigcup_{\sum_{k=1}^\ell} g^k, \forall g^k \in \mathbb{G}$.  \textcolor{black}{($G$ is a virtual market)}.

\STATE Run McAfee auction to select winners, 
denoted by two sets:
$\mathcal{B}^c$, buyer set;
and $\mathcal{S}^c$, seller set. 
McAfee auction also determines the payment $p_{i}^{mcafee}$ for each user $u_i$. 

\STATE Run Group Matching algorithm (Algorithm \ref{alg.gm}) 
with inputs $\mathcal{B}^c$ and $\mathcal{S}^c$ and $\mathbb{G}$,
and obtain a set of final winning buyers $\mathcal{B}^W$ and a set of final winning sellers $\mathcal{S}^W$. 
Determine the final payment: for buyer $u^b_i \in \mathcal{B}^W $, $p_i=\max\{p_{i}^{mcafee}, q_i\}$;
for seller $u^s_j \in \mathcal{S}^W $, $p_j=\min\{p_{j}^{mcafee}, q_j\}$, where $q_i, q_j$
are admission prices. 

\end{algorithmic}
\end{algorithm}

\subsubsection{Group matching algorithm}\label{sec_group_matching}

Group Matching Algorithm, displayed in Algorithm \ref{alg.gm}, chooses the final winning buyers
and sellers from $\mathcal{B}^c$ and $\mathcal{S}^c$.
The idea of this algorithm is as follows.
We first build a bipartite graph $(V,E)$ by connecting a buyer and a seller with an edge if they belong to 
at least one common group. Then, use the enumeration algorithm in \cite{uno97} to find all maximum matchings
in the bipartite graph. Let $\mathcal{M}$ denote the set of all those matchings. 
If $\mathcal{M}=1$, we choose all those vertices (\textit{i.e.}, users) in the unique matching in $\mathcal{M}$
as winners in $\mathcal{B}^W$ and $\mathcal{S}^W$. If $\mathcal{M}>1$,
we first put all those vertices that are covered by all matchings in $\mathcal{M}$ into 
$\mathcal{B}^W$ and $\mathcal{S}^W$. Then, sort other vertices
in descending order of node degree to get a sorted list $U'_r=(u'_1, u'_2, ..., u'_{|U'_r|})$. 
We do the following loop by looking at the nodes one by one in the sorted list,
starting from the first node $u_1'$ (highest degree node):
when looking at $u'_i$, remove those matchings (from the remaining matchings in $\mathcal{M}$)
which do not cover $u'_i$. 
The loop stops when having looked at all nodes or $\mathcal{M}=1$.
This algorithm guarantees that a higher-degree node gets higher priority of being selected
as a final winner. Thus, the dominant strategy of a user is to report as many groups (where she is actually in) as possible,
\textit{i.e.}, report her true group membership information. 
Thus, we have the following lemma. 
\begin{lemma}
With Group Matching algorithm, a user's dominant strategy is to report her true group membership. 
\label{lem.gm}
\end{lemma}

\begin{algorithm}
\caption{ \textbf{Group Matching Algorithm}}
\label{alg.gm}
\begin{algorithmic}[1]

\REQUIRE $\mathcal{B}^c$, $\mathcal{S}^c$, and 
$\mathbb{G}$ (set of all groups).
 
\STATE From $\mathcal{B}^c$, $\mathcal{S}^c$, and $\mathbb{G}$, build a bipartite graph $(V,E)$,
the vertex set is partitioned into $V=\mathcal{B}^c \bigcup \mathcal{S}^c$.
The edge set $E$ contains an edge between a seller $s^c$ in $\mathcal{S}^c$ and a buyer $b^c$ in $\mathcal{B}^c$ 
if and only if there is at least one group to which they both belong according to their reports. 

\STATE Use the enumerating algorithm in \cite{uno97} to find all maximum matchings in this graph.

\IF{there is only one maximum matching $M$,} 
	\STATE Let $\mathcal{B}^M$ and $\mathcal{S}^M$ respectively denote the sets of 
	matched buyers and sellers. Set $\mathcal{B}^W\leftarrow \mathcal{B}^M$, and $\mathcal{S}^W\leftarrow \mathcal{S}^M$.

\ELSIF{there are multiple maximum matchings}
		
	\STATE Let $\mathcal{M}= \{M_j, j=1, 2, ..., m\}$ denote the collection
			of all those matchings.
			Note that each matching $M_j$ is a set of edges, each with end vertices corresponding 
			to a buyer and a seller.
			
	\STATE Derive set $M_{union}=\bigcup_{j=1}^m M_j$. \\
		$ B_{union}=\{ u^b_k, \mbox{if } \exists  (u^b_k, u^s_i) \in M_{union} \}$, \\
		$ S_{union}=\{ u^s_k, \mbox{if } \exists (u^b_i, u^s_k) \in M_{union} \}$.
	
	\STATE Derive set $M_{inter}=\bigcap_{j=1}^m M_j$.
		Let $B_{inter}$ and $S_{inter}$ denote the buyers and sellers in $M_{inter}$ respectively. 
	
	\STATE $\mathcal{B}^W \leftarrow B_{inter}$ and $\mathcal{S}^W \leftarrow S_{inter}$.
	
	\STATE $B'\leftarrow B_{union} \setminus B_{inter}$ and $S'\leftarrow S_{union} \setminus S_{inter}$.\\ 
			$U'=B'\bigcup S'$
	\STATE Sort all users in $U'$ in descending order of node degree to get $U'_r=(u'_1, u'_2, ..., u'_{|U'_r|})$,
	with ties broken arbitrarily.
	
	\FOR{$i=1$ to $|U'_r|$} 
			\STATE If $|\mathcal{M}|=1$, break. 	
		\FOR{$j=1$ \TO $|\mathcal{M}|$ } 
			\STATE If $|\mathcal{M}|=1$, break. 
			\STATE Else if $u'_i\notin M_j$, remove $M_j$ from $\mathcal{M}$.
		
		\ENDFOR
	\ENDFOR
	
	\STATE If $|\mathcal{M}|=1$, add all buyers and sellers in the only matching in $\mathcal{M}$ 
	into sets $\mathcal{B}^W, \mathcal{S}^W$.
	If  $|\mathcal{M}|>1$, arbitrarily choose one matching in $\mathcal{M}$, 
	then add all buyers and sellers in it into sets $\mathcal{B}^W, \mathcal{S}^W$.
	
\ENDIF
\RETURN $\mathcal{B}^W, \mathcal{S}^W$
\end{algorithmic}
\end{algorithm}

\begin{figure}[htb!]
\centerline{
    \begin{minipage}{1.5in}
      \begin{center}
        \setlength{\epsfxsize}{1.4in}
\epsffile{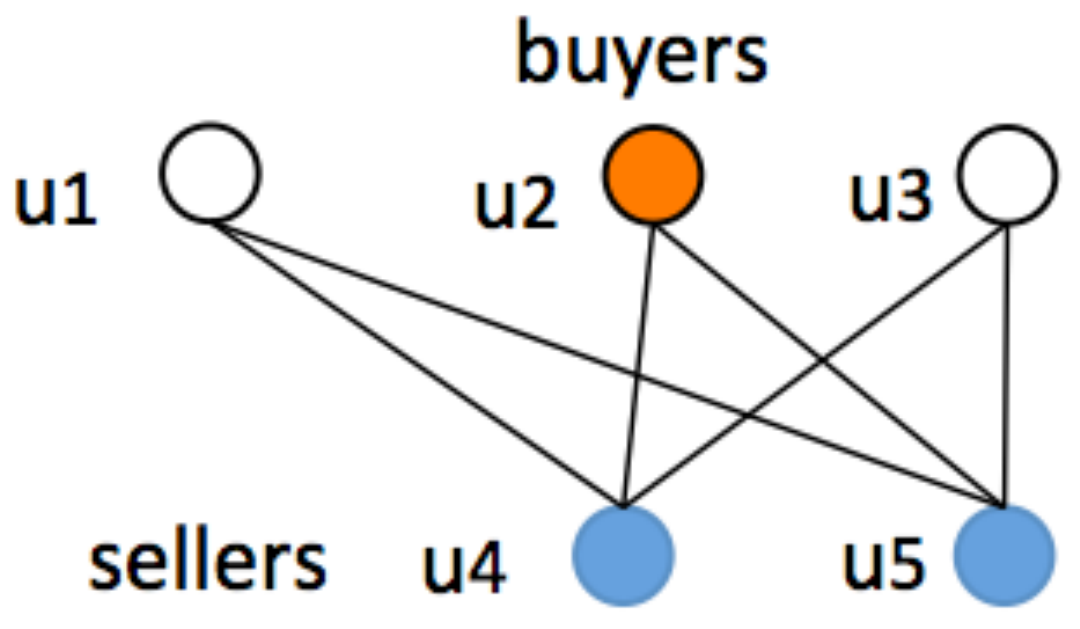}\\
       {}
      \end{center}
\caption{Users report their true group membership.}
\label{fig.ex01}
    \end{minipage}  
    \hspace{4pt} 
    \begin{minipage}{1.5in}
      \begin{center}
        \setlength{\epsfxsize}{1.4in}
\epsffile{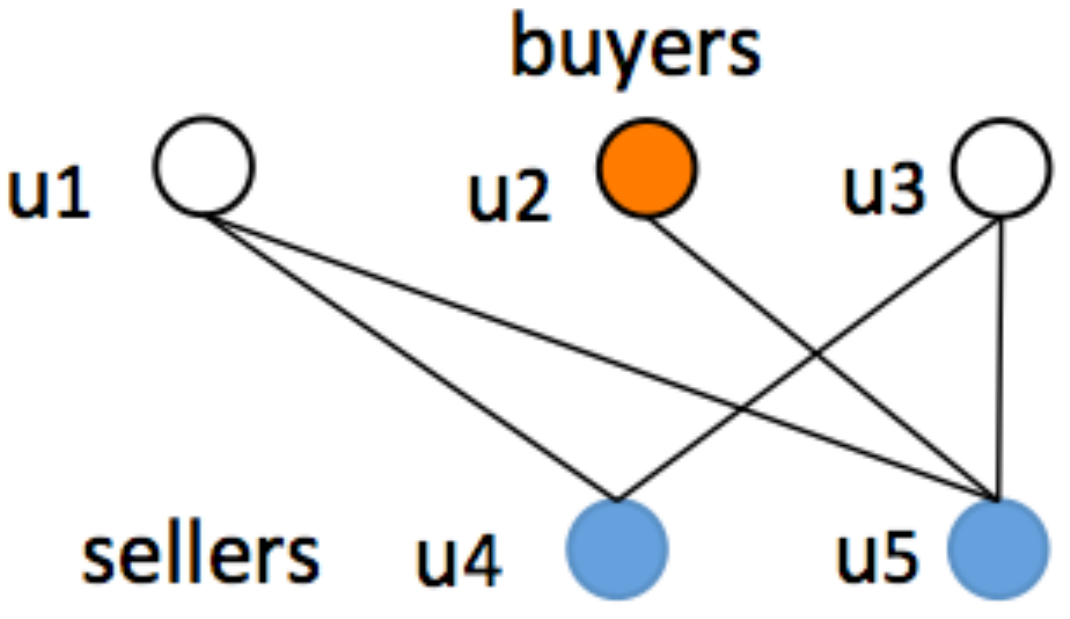}\\
       {}
      \end{center}
\caption{User $u_2$ reports that she is not in the same group as $u_4$.}
\label{fig.ex02}
    \end{minipage}  
    \hspace{4pt} 
}
\end{figure}

Group Matching algorithm discourages a user from cheating about her group membership (\textit{i.e.}, reporting
fewer groups).
An example is given in Figures \ref{fig.ex01} and \ref{fig.ex02}, 
where buyers are $u_1, u_2$, and $u_3$; and sellers are $u_4$ and $u_5$. An edge between a buyer and a 
seller indicates that they are in at least one common group. 
As shown in Figure \ref{fig.ex01}, if user $u_2$ truthfully reports that she is in a group with $u_4$
and $u_5$, then she can possibly be chosen by Group Matching algorithm, as her node degree is the same as that of 
$u_1$ and $u_3$ (note that $u_1, u_2, u_3$ are in $S'$). 
As shown in Figure \ref{fig.ex02}, if $u_2$ sends a mis-report saying that she is only in a group with $u_5$,
then she will certainly not be chosen by Group Matching algorithm.  

VM matching rule is computationally efficient. 
Note that the sorting in McAfee auction  (Step 2 of Algorithm \ref{alg.GPM}) takes polynomial time.
Now consider the computationally dominant parts of Group Matching algorithm (Algorithm \ref{alg.gm}).
The bipartite graph formation in Step 1 takes $O(|E|)$ time. 
The maximum-matching enumeration takes $O(|V||\mathcal{M}|)$ time with $(|V|=|\mathcal{B}^c|+|\mathcal{S}^c|)$.
This is because the enumeration algorithm in \cite{uno97} takes $O(|V|)$ time per max-matching.
Steps 3-10 take $O(|E|^2 |\mathcal{M}|)$ time. 
The sorting in Step 11 takes $O(|V|\log(|V|)$ time.
The nested loops (from Step 12 to Step 18) take $O(|V||\mathcal{M}|)$ time. 
Note that if the graph is sparse and/or small (which is usually the case), 
$|\mathcal{M}|$ is a small number.
Empirical data from the MIT Reality Mining 
project \cite{eagle2006reality} indicates that this is indeed the case in practice
(See Section \ref{sec_mit}).
Thus this algorithm is practically efficient.
But if the bipartite graph is dense and large, $|\mathcal{M}|$ can be very large. 
In that case,
we use a truthful polynomial-time heuristic to determine $\mathcal{B}^W$ and $\mathcal{S}^W$.
The basic idea is that we start with the bipartite graph, 
and sort all buyer nodes and all seller nodes in descending order of node degree
to get two sorted list $U'_B=(u^b_{(1)}, u^b_{(2)},...,u^b_{(|U'_B|)})$ (buyers) 
and  $U'_S=(u^s_{(1)}, u^s_{(2)},...,u^s_{(|U'_S|)})$ (sellers). 
Out of $u^b_{(1)}$ and $u^s_{(1)}$, choose the one with higher degree (with ties broken arbitrarily).
Suppose that $u^b_{(1)}$ is chosen. 
Then search $U'_S$ from the beginning till find the highest-degree seller node $u^s_{(k)}$
that is in a common group with $u^b_{(1)}$, and then put $u^b_{(1)}$ and $u^s_{(k)}$ 
into $\mathcal{B}^W$ and $\mathcal{S}^W$ respectively,
and remove them from the graph and from their respective lists. 
If such a $u^s_{(k)}$ does not exist, remove $u^b_{(1)}$ from $U'_B$ and the graph.
Continue this process till either $U'_B=\emptyset$, or $U'_S=\emptyset$, or $E=\emptyset$.  
This heuristic clearly takes polynomial time. When compared with Algorithm \ref{alg.gm},
this heuristic might not return the maximum number of buyer/seller pairs (hence a performance loss). 
However, our simulations show that the difference between it and Algorithm \ref{alg.gm}
is very small. This heuristic is truthful, as a higher-degree node gets higher priority of being selected.

\subsubsection{Strong no-trade construction}

We borrow the SNT construction from \cite{bredin07chain}, shown below.
\begin{definition}
$SNT(t)$ set 
is defined as
$SNT(t):=\emptyset$ if $\min(|\mathcal{B}(t)|, |\mathcal{S}(t)|)\ge 2$; 
otherwise $SNT(t):=\mathcal{B}(t) \bigcup \mathcal{S}(t)$.
\end{definition}\label{def_SNT}
This is a valid SNT construction because an empty set is always a valid SNT construction, 
and no user will be selected for trade according
to the VM matching rule when $\min(|\mathcal{B}(t)|, |\mathcal{S}(t)|)< 2$, thus this construction is valid.

\subsubsection{An example}

\begin{figure}[htb!]
\centerline{
    \begin{minipage}{1.5in}
      \begin{center}
        \setlength{\epsfxsize}{1.3in}
\epsffile{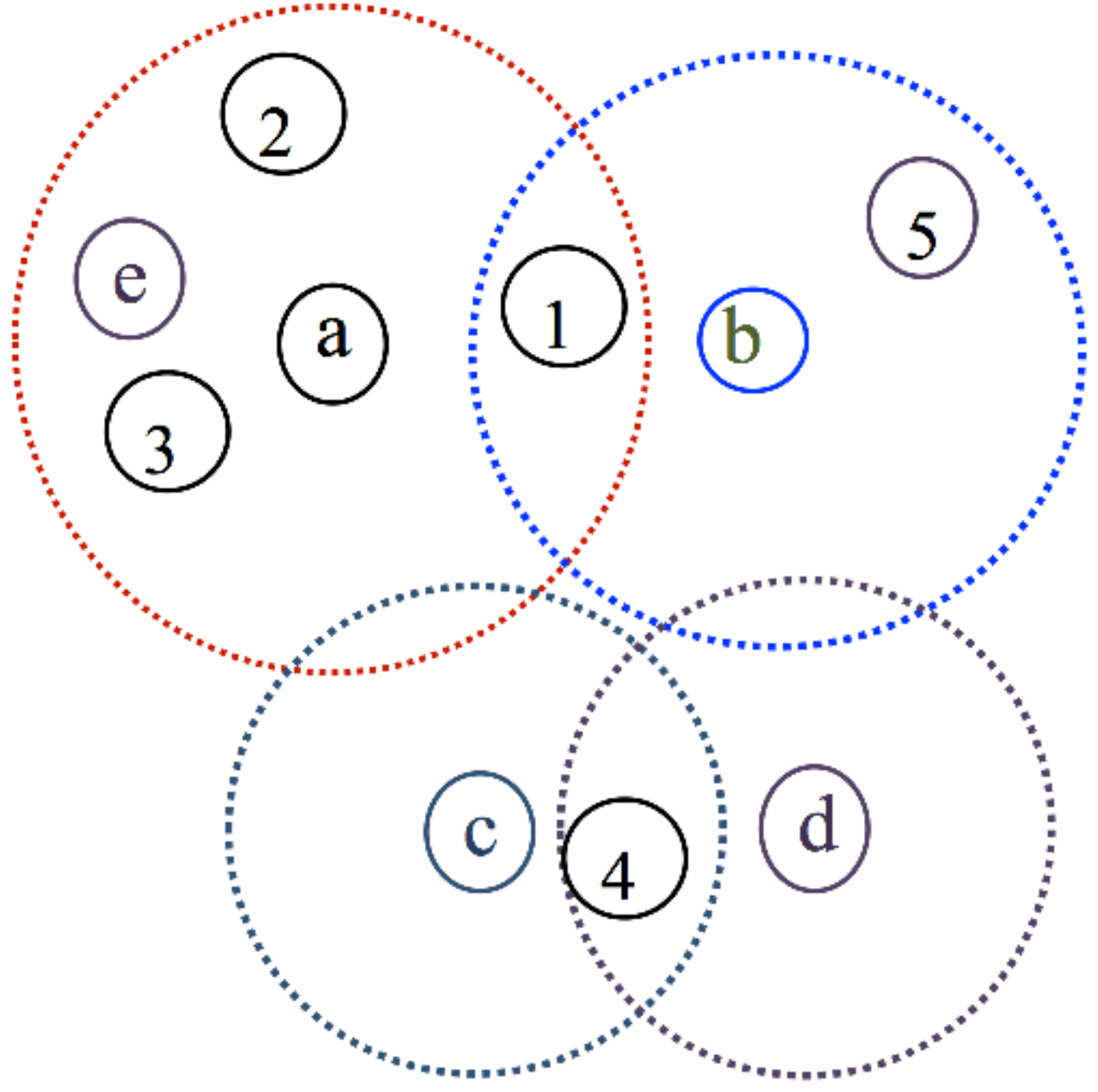}\\       
	{}
      \end{center}
\caption{Example set of active users.}
\label{fig.toy1}
    \end{minipage}  
    \hspace{4pt} 
    \begin{minipage}{1.5in}
      \begin{center}
        \setlength{\epsfxsize}{1.4in}
\epsffile{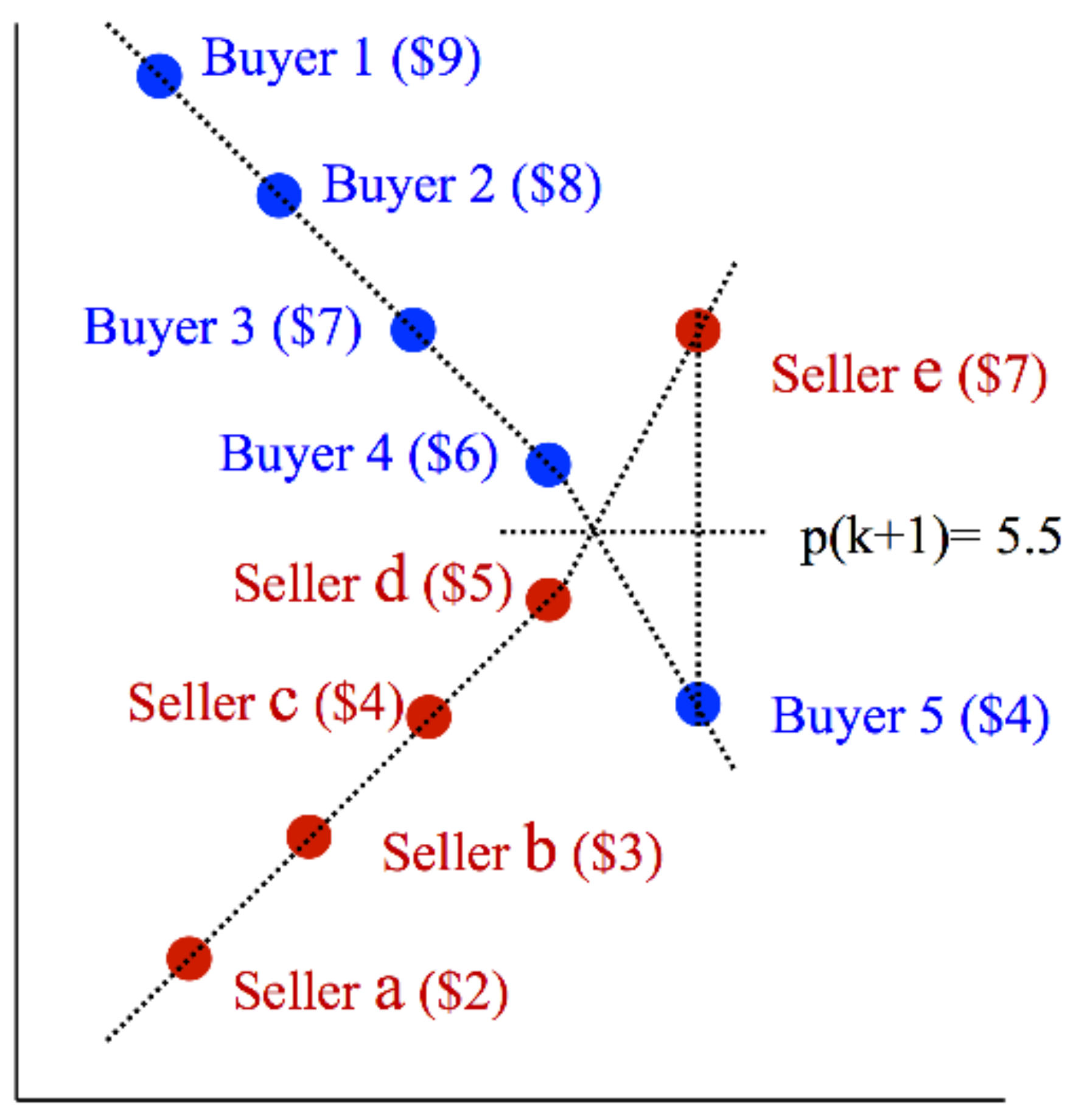}\\
       {}
      \end{center}
\caption{Step 2 of Algorithm \ref{alg.GPM}.}
\label{fig.toy2}
    \end{minipage}  
    \hspace{4pt} 
}
\end{figure}

\begin{figure}[htb!]
\centerline{
    \begin{minipage}{1.5in}
      \begin{center}
        \setlength{\epsfxsize}{0.5in}
\epsffile{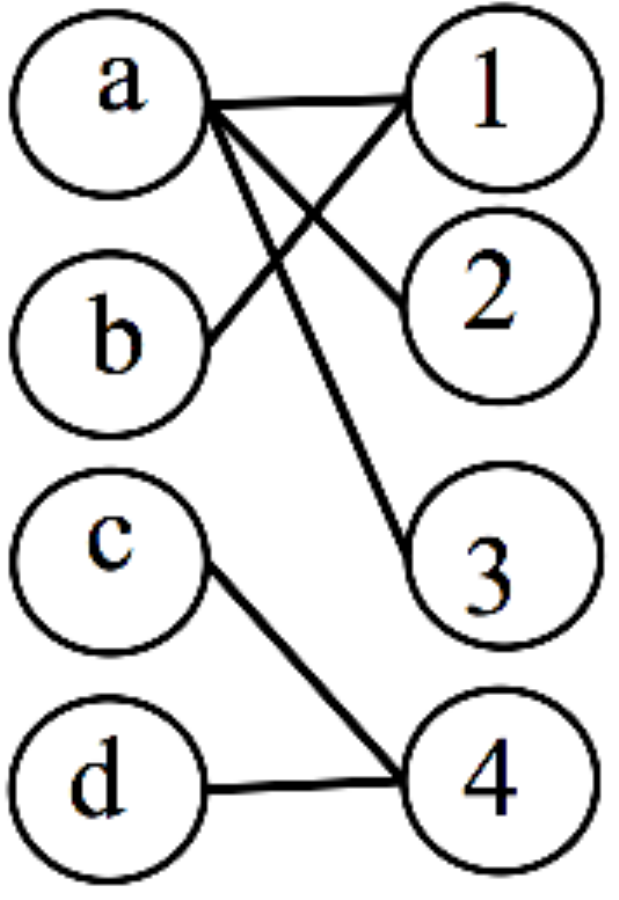}\\
       {}
      \end{center}
\caption{The bipartite graph at the beginning of Algorithm \ref{alg.gm}.}
\label{fig.toy3}
    \end{minipage}  
    \hspace{4pt} 
    \begin{minipage}{1.5in}
      \begin{center}
        \setlength{\epsfxsize}{0.5in}
\epsffile{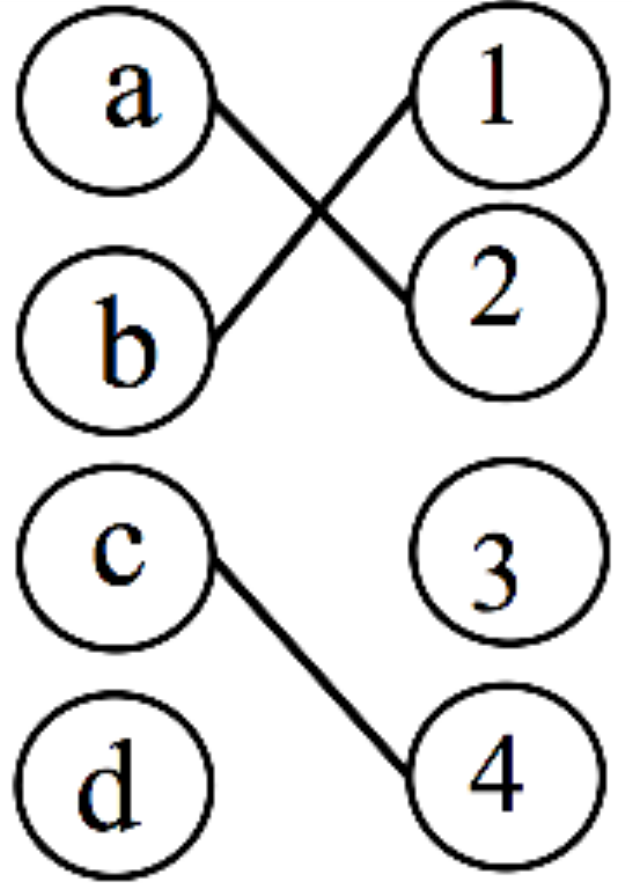}\\
       {}
      \end{center}
\caption{The final winning matching set $M_4$ at the end of Algorithm \ref{alg.gm}.}
\label{fig.toy4}
    \end{minipage}  
    \hspace{4pt} 
}
\end{figure}

\textcolor{black}{
We now illustrate Algorithms \ref{alg.GPM} and \ref{alg.gm} through the following example. 
Consider a network topology with a set of active buyers $\mathcal{B}=\{1,2,3,4,5\}$ 
and a set of active sellers $\mathcal{S}=\{a,b,c,d,e\}$ 
at a particular time period, as depicted in Figure \ref{fig.toy1}.
The candidate selection process (Step 2 in Algorithm \ref{alg.GPM}) is demonstrated in Figure \ref{fig.toy2}.  
The algorithm selects $4$ candidate sellers (out of $5$) 
and $4$ candidate buyers (out of $5$), by placing all users
in a virtual market or group. 
Then at the beginning of Algorithm \ref{alg.gm}, 
we construct a bipartite graph (shown in Figure \ref{fig.toy3})
and find its set of all max matchings  
$\mathcal{M}=\{ M_1=\{(a, 3), (b, 1), (c, 4)\},
M_2=\{(a, 3), (b, 1), (d, 4)\},
M_3=\{(a, 2), (b, 1), (d, 4)\},\\
M_4=\{(a, 2), (b, 1), (c, 4)\} \}$.
Then in Steps $8$ and $9$ of Algorithm \ref{alg.gm}, we have $M_{inter}= \{(b, 1) \}$, 
$\mathcal{B}^W=\{ 1\}$, and $\mathcal{S}^W=\{b \}$.
After Steps $10$ and $11$, the sorted list is $U'_r = (a, 4, 2, 3, c, d )$.
Then the loop during Steps $12$ to $18$ includes the following four iterations:
Iter-1, $\mathcal{M}$ remains unchanged;
Iter-2, $ \mathcal{M}$ remains unchanged;
Iter-3, $ \mathcal{M}=\{ M_3, M_4 \}$; 
Iter-4, $\mathcal{M}=\{ M_4 \}$.
Finally in Step 19, we have  
$\mathcal{S}^W=\{a, b, c\}$ and $\mathcal{B}^W =\{ 1, 2, 4\}$
as final winners, as depicted in Figure \ref{fig.toy4}.
}

\subsection{Truthfulness and computational efficiency of M-CHAIN}

We prove that VM matching rule is a well-defined single-period matching rule.
\textcolor{black}{See Appendix for a proof. } 

\begin{theorem}
Virtual Market single-period matching rule is well-defined, \textit{i.e.}, it is truthful, no-deficit, individual-rational, and feasible.
\label{them.GPM}
\end{theorem}

M-CHAIN satisfies all requirements for an DDA CHAIN auction.  
Based on Lemma \ref{lemma_da_chain} (Theorem 3 in \cite{bredin07chain}), we have:
\begin{theorem}
Multi-market dynamic double-auction\\ M-CHAIN is truthful, no-deficit, feasible, and individual-rational.
\end{theorem}

M-CHAIN auction is computationally efficient, because  
the Virtual Market matching rule (the computationally dominant part of M-CHAIN) is efficient.

\subsubsection{Variants of M-CHAIN} \label{sec_extent}

M-CHAIN can also adopt other well-defined single-period matching rules designed for a multi-market system. 
For example, we have designed a 
random ordering multi-market (ROM) rule that puts all groups (in period $t$) into a random order,
then treats the ordered list of groups as an instance of a single-group dynamic double auction.  
Due to space limitations, we will defer the discussion of this rule in an extended version of this paper.

\subsubsection{Heterogeneous services} \label{sec_hetg}

\textcolor{black}{We next consider heterogeneous service where the level of quality of service offered by sellers is not uniform, 
e.g., two sellers may offer data connections with different signal strengths. 
There might be many possible design goals for a heterogeneous system. 
One practical goal we consider here is to design an auction algorithm that selects a set of buyers and sellers 
that achieves the maximum overall quality of service (with ties broken arbitrarily), subject to 
the requirement that the algorithm must be truthful, individually rational, feasible, and no-deficit. 
We can slightly modify our M-CHAIN algorithm to achieve this design goal. Specifically, 
we can assign a positive weight value to every edge in the bipartite graph $(V,E)$ 
in the initial step of the Group Matching algorithm (Algorithm \ref{alg.gm}).
Those edge weights can be set proportional to the levels of quality of service offered by sellers.
Then, at Step $6$, we compute the sum of all edge weights  
in each matching set $M_j, j=1,2,\ldots,m$ and 
sort those matchings in a descending order according to their sums of edge weights. 
If there is only one matching with maximum sum of edge weights, denoted by $M_{max}$, then 
let $\mathcal{M}=\{M_{max}\}$ and 
the algorithm jumps to Step 19; if there are multiple matchings with the same
maximum sum of edge weights, let $\mathcal{M}_{max}$ denote the set of those matchings, and 
let $\mathcal{M}=\mathcal{M}_{max}$, and then the algorithm continues at Step $7$. 
This modified algorithm guarantees the truthfulness from users and ensures 
the set of finally chosen buyers and sellers achieves maximum quality of service
among all candidate buyers/sellers.}

\section{Performance Evaluation} \label{sec_evaluation}

In this section, we evaluate the system efficiency and the price of truthfulness guarantee of
M-CHAIN auction by simulating users' patterns
based on generated data and a real-world dataset.
Due to space limitations, we focus on allocative efficiency, a typical efficiency metric. 

\subsection{Evaluation metrics}

\subsubsection{Efficiency}

We have developed a Python-based simulator to simulate DMACS systems, each characterized by 
a tuple  $$\langle \mathbb{B}(T), \mathbb{S}(T), \mathbb{G}(T), T, K\rangle,$$ 
\textit{i.e.}, a problem instance that is solved by M-CHAIN.
For each simulation run of a problem instance, we aggregate 
the total value (denoted by $V_{mChain}$) of all users that are selected for trade during all time periods. Formally,
$
V_{mChain} = \sum_{t\in T, u^b_i\in \mathcal{B}^W(t),u^s_i\in \mathcal{S}^W(t)} (b_i-s_j).
$

For each simulation run, we also calculate the total value of an offline optimal matching,
which is found by solving the following integer programming problem,
based on the assumption that we know all users' true types.
\begin{eqnarray}
 \max_{x_{ij}} & \sum_{(u^b_i, u^s_j)\in F} x_{ij}(b_i - s_j) &  \label{eqn.opt}\\
 s. t. & 0\le  \sum_{i: (u^b_i, u^s_j)\in F } x_{ij}  \le  1,     & \forall u^s_j \\ 
 & 0\le  \sum_{j: (u^b_i, u^s_j)\in F } x_{ij}  \le  1,     & \forall u^b_i \\ 
 & x_{ij} \in \{0,1\}, & \forall u^b_i, u^s_j 
\end{eqnarray} 
Set $F$ consists of all pairs of buyer and seller that can appear in the same period and in the same group. 
\begin{eqnarray}
F = \{(u^b_i, u^s_j) &:& a_i \le d_j, a_j \le d_i, \nonumber \\
&& \exists t, k, \mbox{ s.t. } u^b_i \in g^k(t), u^s_j \in g^k(t)\}
\end{eqnarray}
Let $V_{opt}$ denote the total value derived by solving the integer programming problem in (\ref{eqn.opt}).
Then a normalized allocative efficiency of M-CHAIN solving a DMACS problem instance is defined as
$E = V_{mChain} / V_{opt}$. For the rest of the paper, we will call $E$ \textit{efficiency}.
We run each simulation setting with $5$ different random seeds, and calculate their average efficiency 
as M-CHAIN's efficiency in that setting.

\subsubsection{Price of truthfulness guarantee}
We are also interested in the amount of efficiency loss due to the truthfulness guarantee of 
an auction design, referred to as \textit{price of truthfulness guarantee}.
Recall that Myerson and Satterthwaite \cite{myerson1983efficient} shows that no double auction 
can simultaneously achieve system efficiency, truthfulness, individual rationality, and no-deficit. 
We investigate the {price of truthfulness guarantee} of M-CHAIN by 
comparing its efficiency level with that of a \textit{random greedy online matching} algorithm. 
The random greedy algorithm works as follows.
In each period $t$, we first put all groups into a randomly ordered list. 
Then we conduct a single group greedy matching for each group, 
following the order of the list. A user that is not selected in any group's greedy matching
will still stay in the system if her departure time is later than $t$.
The single group greedy matching (from \cite{bredin07chain}) 
is done by first ordering buyers and sellers into two lists (as what McAfee auction \cite{mcafee1992dominant} initially does),
then select pairs of buyers and sellers till the first pair in which the buyer's value is no more than 
the seller's value.
\textit{Note that this is not a truthful algorithm.}
It would achieve a higher efficiency level than that of M-CHAIN
if all users were truthful. Let $V_g$ denote the total value of the random greedy algorithm.
We define M-CHAIN's price of truthfulness guarantee as $L_{mChain}=(V_{g}-V_{mChain}) / V_{opt}$.

\subsection{Stochastic arrivals and random grouping of users}

In each simulation run, users arrive in the system following a Poisson arrival process, 
with \textit{mean inter-arrival time} varying between $0.1$ and $1.5$ time periods. 
In addition, peers' maximum patience is $K=6$. 
Each user's patience is taken uniformly at random from $[0,6]$.
For each period $t$, there are at most $10$ groups, and out of which, a user present
in the system can be in randomly chosen $\ell$ groups, with $\ell\in [1,10]$. 
Each user's value is taken from a uniform random distribution with lower and upper bounds of it
are $20\%$ less than or more than its mean, and the mean varies over time following 
a Brownian motion. Specifically, in each $t$, we randomly increase or decrease the mean 
by multiplying it with $e^{\pm\gamma}$, where $\gamma$ is a market volatility factor varying from 
$0.01$ to $0.15$. The initial mean value for a problem instance is $20$.
The total number of users arriving at the system is $10,000$, and when a new user joins the system, she is
equally likely to be buyer or a seller. 
The arrival process and market volatility patten are similar to those in \cite{bredin07chain}, 
but \cite{bredin07chain} does not consider multiple groups. 
We next present a few typical results from our extensive simulations. 
 
\subsubsection{Efficiency varies as a function of arrival rate and market volatility}\label{sec_effc_poisson}

\begin{figure}[htb!]
\centerline{
    \begin{minipage}{1.6in}
      \begin{center}
        \setlength{\epsfxsize}{1.6in}
\epsffile{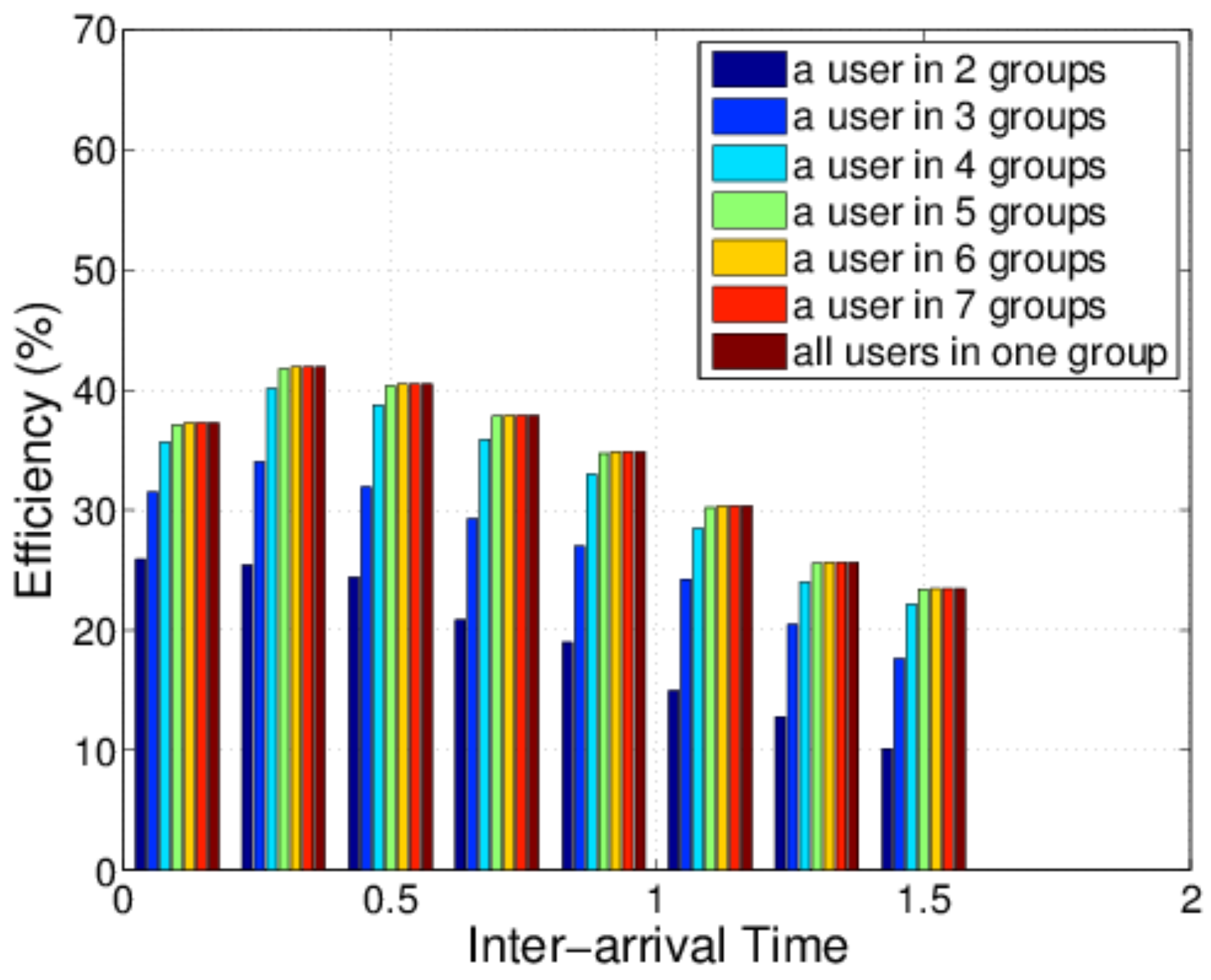}\\
       {}
      \end{center}
\caption{Efficiency comparison between $7$ cases when market volatility $\gamma=0.01$.}
\label{fig.sim_effc_GM_10_SG_0_PTN_6_MVI_20_AgentsLm_5000_GrpMatch_1_volatility_0.01}
    \end{minipage} 
      \hspace{4pt} 
    \begin{minipage}{1.6in}
      \begin{center}
        \setlength{\epsfxsize}{1.6in}
\epsffile{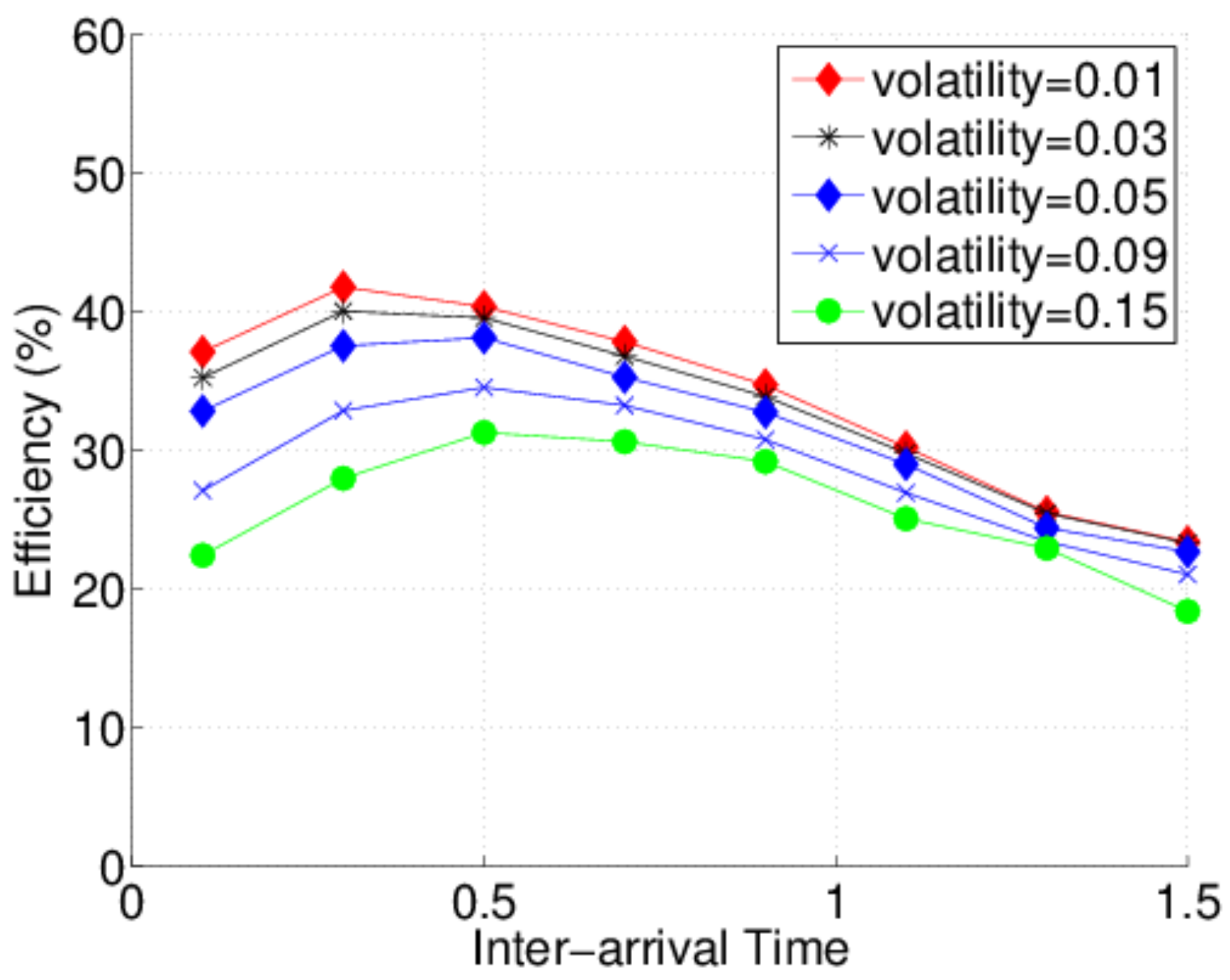}\\
       {}
      \end{center}
\caption{Efficiency comparison.
A user is in 5 randomly chosen groups (out of 10) in each period. }
\label{fig.sim_effc_GM_10_SG_1_PTN_6_MVI_20_AgentsLm_10000_GrpMatch_2_Grp_5}
    \end{minipage}  
}
\end{figure}

Figure \ref{fig.sim_effc_GM_10_SG_0_PTN_6_MVI_20_AgentsLm_5000_GrpMatch_1_volatility_0.01} 
compares the cases where each user is in $2, 3, 4, 5$, $6$, and $7$ randomly chosen groups, and 
all users in a single group (when $\gamma=0.01$). We observe that M-CHAIN's efficiency $E$ increases
as a user belongs to a larger number (denoted by $\ell$) of randomly chosen groups. 
This is because a cluster of overlapping multiple groups will approach a single unified group
as $\ell$ increases, and this figure shows that when $\ell=5,6, 7$, M-CHAIN's efficiency level
is almost the same as that of a single-market case. We observe similar pattern
in all other volatility levels. 
Figure \ref{fig.sim_effc_GM_10_SG_1_PTN_6_MVI_20_AgentsLm_10000_GrpMatch_2_Grp_5}
shows the impact of market volatility factor $\gamma$ on M-CHAIN's efficiency. We observe that 
as market becomes more volatile (\textit{i.e.}, larger $\gamma$), the efficiency becomes lower,
which is consistent with our intuition.
In addition, except for the increase at the beginning,
the efficiency decreases in general as the inter-arrival time increases, because 
as users join the system less frequently, they are more likely distributed into separated clusters.
The initial increase in those curves can be explained by our stringent SNT construction
which can eliminate lots of users from the auction if there is a large number of users in system.
Part of our ongoing research is to develop a formal analysis of this pattern.  
In addition, all those curves (in Figure \ref{fig.sim_effc_GM_10_SG_1_PTN_6_MVI_20_AgentsLm_10000_GrpMatch_2_Grp_5}) 
move closer to each other as inter-arrival time gets larger, because
if users join the system very infrequently, the market volatility won't have much impact
on the efficiency as there are not many users in the system.  

\subsubsection{Price of truthfulness guarantee}

\begin{figure}[htb!]
\centerline{
    \begin{minipage}{1.65in}
      \begin{center}
        \setlength{\epsfxsize}{1.65in}
\epsffile{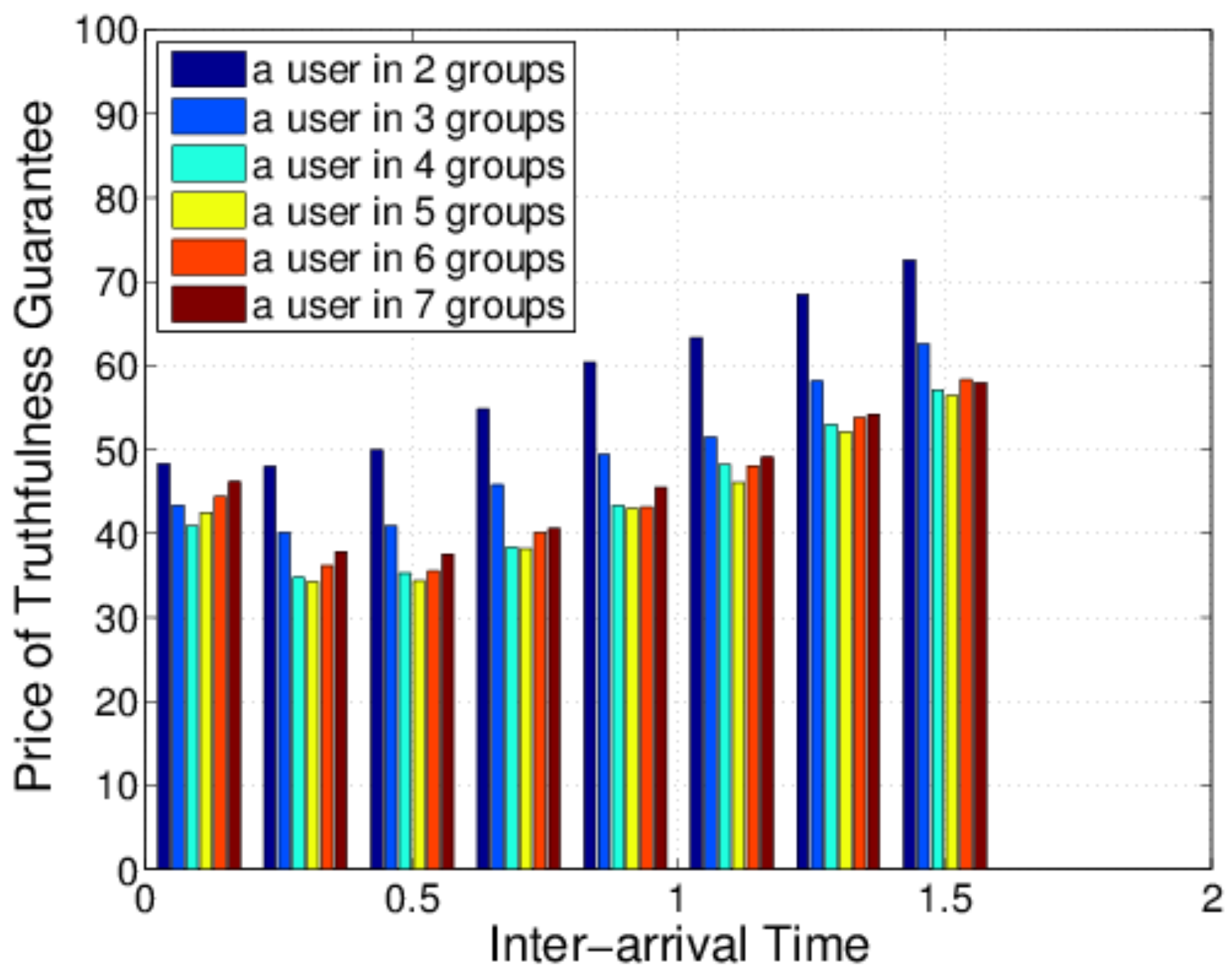}\\
       {}
      \end{center}
    \end{minipage} 
}
\caption{Price of truthfulness guarantee when $\gamma=0.05$.}
\label{fig.sim_effc_GM_10_SG_1_PTN_6_MVI_20_AgentsLm_10000_GrpMatch_2_Loss_Cheat_vol_0.05}
\end{figure}

Figure \ref{fig.sim_effc_GM_10_SG_1_PTN_6_MVI_20_AgentsLm_10000_GrpMatch_2_Loss_Cheat_vol_0.05}
shows the price of truthfulness guarantee of
M-CHAIN when compared with random greedy online matching algorithm
(given that $\gamma=0.05$).
We observe (across all simulation settings) that the efficiency loss can be large in order to guarantee the truthfulness from users,
which is consistent with the impossibility result by Myerson and Satterthwaite \cite{myerson1983efficient} 
that suggests there is a tradeoff between truthfulness and efficiency.
Our results here further indicate a high price we need to pay for guaranteed truthfulness
in a multi-market dynamic double auction setting. 

\parskip=-5pt
\subsection{
M-CHAIN in a highly dynamic and sparse system}\label{sec_mit}
\parskip=2pt

We now evaluate M-CHAIN based on the user patterns 
(including user arrival/departure processes and groupings)
derived from a real-world dataset from the MIT Reality Mining project \cite{eagle2006reality}. 
The information in the dataset includes call logs, Bluetooth
devices in proximity, etc., that were collected from 
$100$ users (from August $2004$ to July $2005$) who carried Nokia 6600 smartphones.
We are only interested in the device proximity information, 
\textit{i.e.}, the records of each user's neighbors that were detected via every $5$-min Bluetooth scan
through their smartphones.
We did not use the data of July $2005$, as that month contains less than $100$ scan records,
whereas other months contain more than thousands of records.  
We first process the raw data and derive group information per scan per day. 
A cluster of groups in a DMACS system corresponds to 
a proximity network \cite{eagle2006reality} 
(\textit{i.e.}, a network of users formed by the Bluetooth-detected proximity information) 
in the dataset.
We choose to interpret each 5-min scan as a single period of a DMACS system.
One might argue that a longer time (e.g., 30 min) might be more reasonable for 
a single period $t$ (e.g., for collaboratively watching a video).
However we are constrained by the existing data.
Our interpretation
is reasonable if we assume that the user pattern
derived from the dataset is similar to what would be observed in a DMACS system. 

\begin{figure}[htb!]
\centerline{
    \begin{minipage}{1.6in}
      \begin{center}
        \setlength{\epsfxsize}{1.6in}
\epsffile{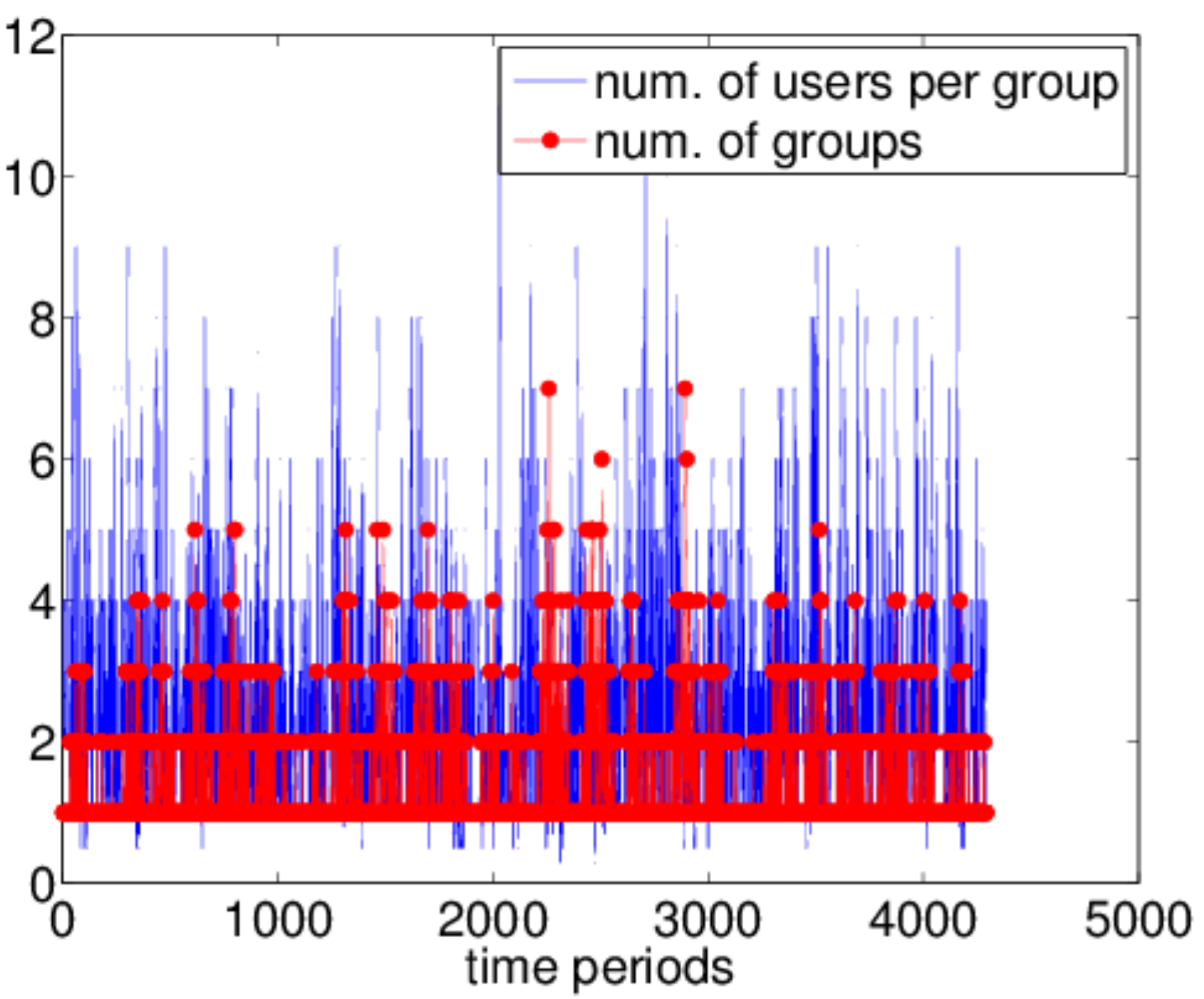}\\
       {}
      \end{center}
\caption{Num. of users per group and num. of groups in Oct. 2014. }
\label{fig.grp_info__CVR_5_CNT_1_200410_series}
    \end{minipage}
    \begin{minipage}{1.7in}
      \begin{center}
        \setlength{\epsfxsize}{1.7in}
\epsffile{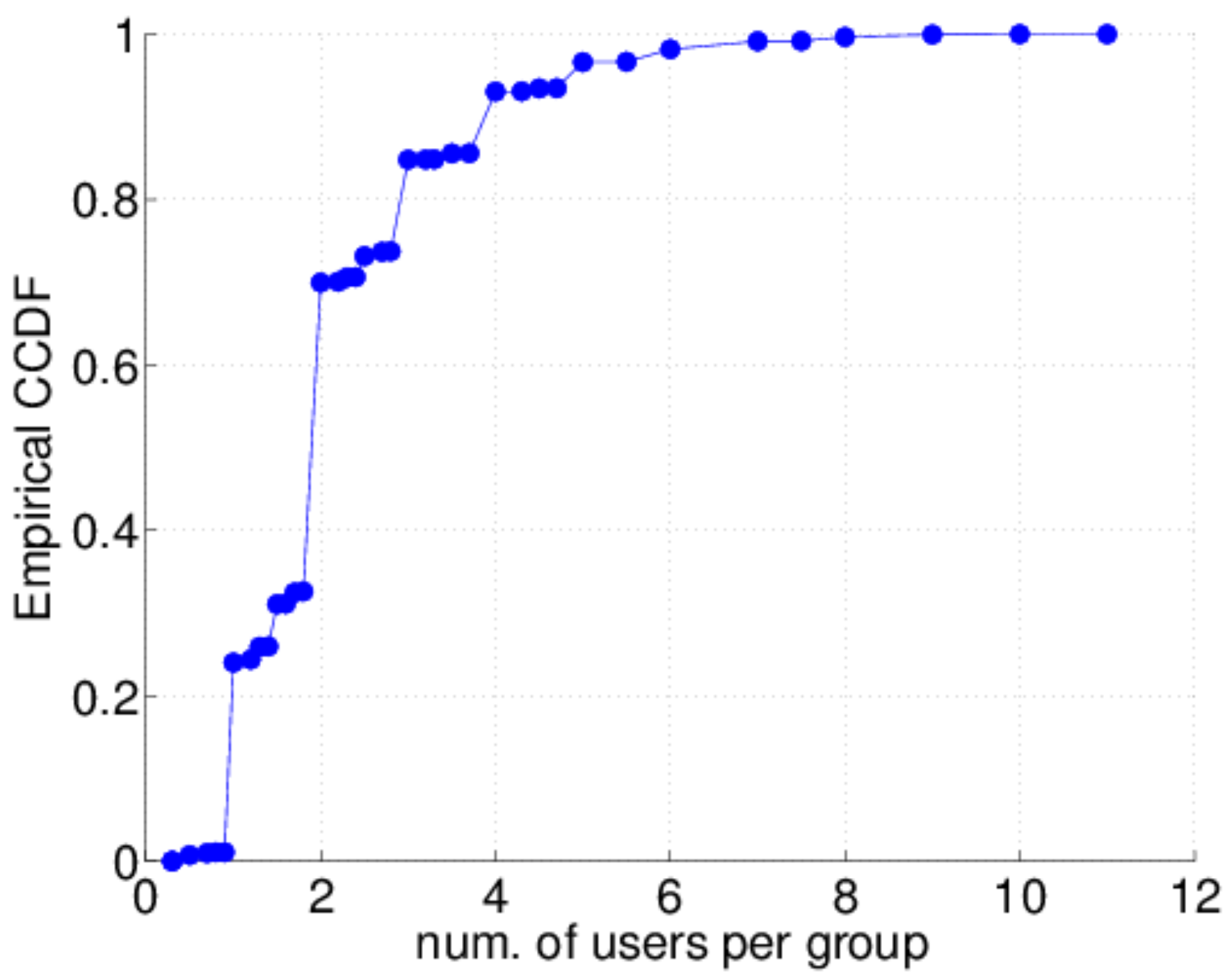}\\
       {}
      \end{center}
\caption{Empirical CDF of num. of users per group in Oct. 2014.  }
\label{fig.grp_info__CVR_5_CNT_1_200410}
    \end{minipage}    
}
\end{figure}

\begin{figure}[htb!]
\centerline{
        \begin{minipage}{1.7in}
          \begin{center}
            \setlength{\epsfxsize}{1.7in}
    \epsffile{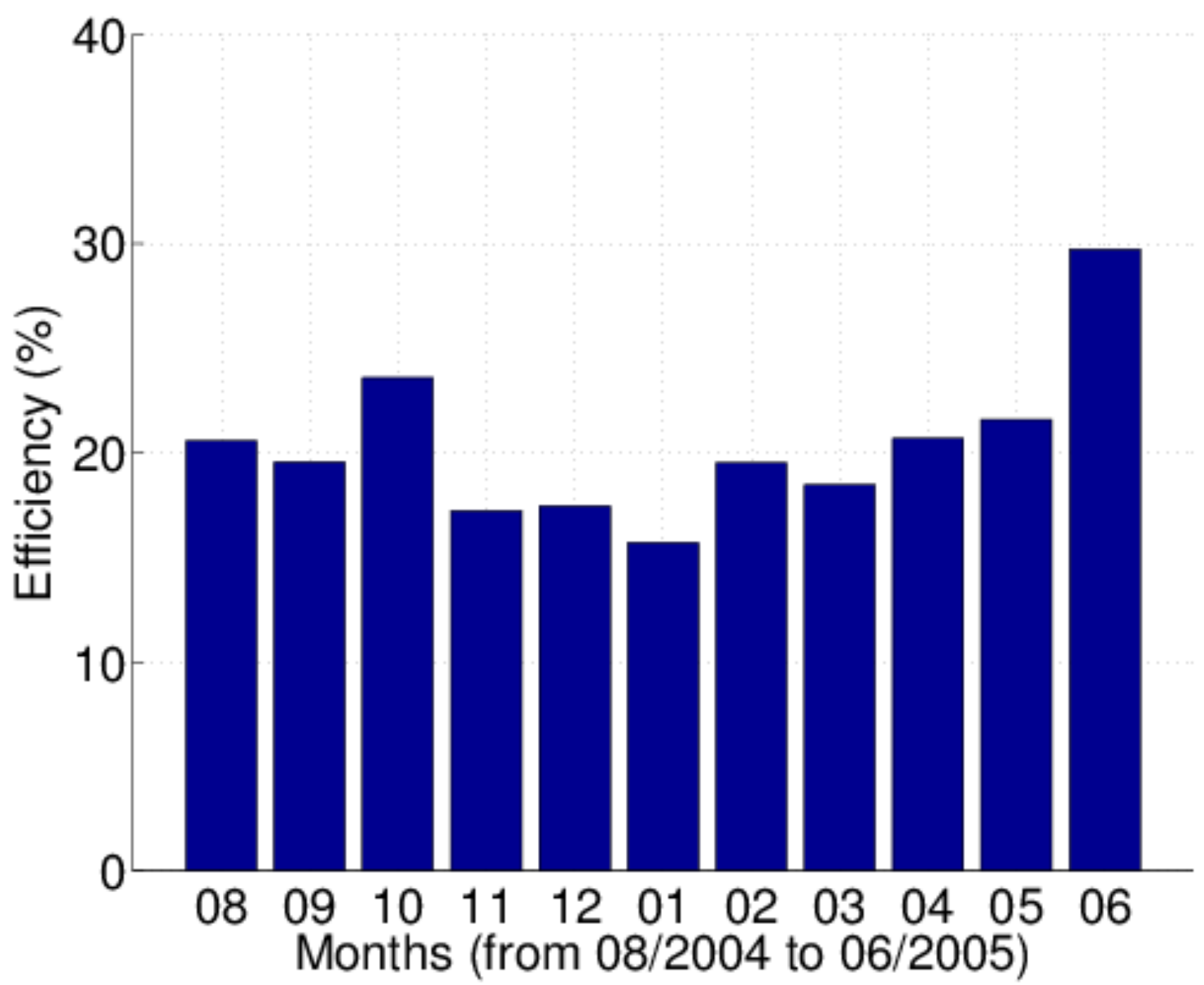}\\
           {}
          \end{center}
        \end{minipage}    
}
    \caption{Efficiency of M-CHAIN in a highly dynamic and sparse system. }
    \label{fig.sim_effc_MIT_CVR_5_CNT_1_months}
\end{figure}

The groups in the dataset are ``extremely dynamic and sparse" \cite{eagle2006reality},
which suggests that at any time instant, a cluster or a group is very unlikely to contain many users,
and the members of each group are changing very frequently.   
Indeed we observe such patterns. For example,  
as shown in Figure \ref{fig.grp_info__CVR_5_CNT_1_200410_series},
the data of October $2004$ shows highly dynamic patterns
of \textit{the number of groups per period} and \textit{the average number of users per group per period}. 
Most of the time, there is only one or two groups in the system. 
Figure \ref{fig.grp_info__CVR_5_CNT_1_200410} further shows that in that month,
about $70\%$ time, there are on average only $2$ or fewer users in a group (for which a trade is very unlikely to happen). 
Thus, we should expect that M-CHAIN has a low efficiency level  in such a dynamic and sparse system.
Nevertheless, Figure \ref{fig.sim_effc_MIT_CVR_5_CNT_1_months} 
shows that M-CHAIN is still able to maintain a reasonable efficiency level (around $20\%$).

\parskip=-5pt
\section{Related Work}\label{sec_related}
\parskip=2pt

Bredin \emph{et al.} \cite{bredin07chain} proposes CHAIN, a design framework for dynamic double auction,
but it only considers a single market. Nevertheless, the techniques in \cite{bredin07chain} give
us some fundamental building blocks for our proposed auction in this paper.  
A pioneering work by Yang \textit{et al.} \cite{yang13location} applies double auction mechanism 
to assist k-anonymity location privacy service, 
but \cite{yang13location} only studies single period
and single-market auction mechanisms. 
Auction-based incentive mechanisms are also recently studied in \cite{xue_mobicomm,koutsopoulos2013optimal}
for crowdsourcing to smartphones,
but their models 
are not double auctions. 
Our work also relies on some of the fundamental works on auction, e.g., McAfee auction \cite{mcafee1992dominant},
Hajiaghayi's online auction \cite{hajiaghayi2005online}, 
and the impossibility result from Myerson and Satterthwaite \cite{myerson1983efficient}. 
Examples of DMACS application include Microcast \cite{keller2012microcast} and
mobile P2P based k-anonymity location privacy \cite{shokri2013hiding}. 

\parskip=-5pt
\section{Conclusion}\label{sec:conclusion}\label{sec_conclusion} 
\parskip=2pt

We designed M-CHAIN, an auction-based incentive mechanism for the emerging user-assisted mobile crowd services, 
referred to as DMACS. 
M-CHAIN is truthful and computationally efficient,
and addresses the multi-market nature of a DMACS system,
which is absent from the existing dynamic double auction designs.  
Our simulations based on generated user patterns
and real-world traces have demonstrated the good efficiency level of M-CHAIN. 
As future work, we will fully implement and test our design
in a real mobile environment.

\input{DMACS_ieee.bbl}

\appendix[Proof of Theorem \ref{them.GPM}]\label{sec_appendix}

Theorem \ref{them.GPM} is re-stated below.

\textit{Virtual Market single-period matching rule is well-defined, that is, 
it is truthful, no-deficit, individual-rational, and feasible.}

\begin{proof} 

Virtual Market single-period matching rule is shown as Algorithm \ref{alg.GPM}.

We first prove that Virtual Market single-period matching rule is truthful.
To that end, we need to prove that both reporting true group membership and 
reporting true value are dominant strategies for a user $u_i$.

Note that in Step 2 of Algorithm \ref{alg.GPM}, 
the McAfee auction does not take into consideration any group membership information. 
Such information is considered in Step 3, the Group Matching algorithm. 
In Step 3, reporting true group membership is a dominant strategy for $u_i$ 
according to Lemma \ref{lem.gm}. Note that this is also true for the heuristic (which is more efficient 
in a dense and large graph than the Group Matching algorithm), as discussed in Section \ref{sec_group_matching}.

Next we show that reporting true value (bid or ask) is a dominant strategy for $u_i$.
We prove this for a buyer only. Similarly we can prove it for a seller. 

Consider buyer $u^b_i$. Her true value is $b_i$, reported bid is $\hat{b}_i$.
We will show that there is no benefit for $u^b_i$ to misreport her bid in the following
two cases. 

Case 1. If $\hat{b}_i = b_i$ and $u^b_i$ is chosen for trade by both Steps 2 and 3 of the Virtual Market rule 
(that is, $u^b_i\in \mathcal{B}^c$ and $u^b_i\in \mathcal{B}^W$),
then there is no benefit to $u^b_i$ if she misreports her bid, that is, $\hat{b}_i \ne b_i$. 
This in because in Step 2 of Algorithm \ref{alg.GPM}, the McAfee auction ensures that the
dominant strategy of a user is to report her true bid. 
The fact that multiple overlapping
groups are pooled together as a single virtual group has no impact on the truthfulness property of
the McAfee auction as it only takes values into consideration when
determining who will be the winners and the amount of payment for them. 
A possible impact of the formation of the virtual group on $u_i$ is that even if she reports
her true value, she might not be selected as one of the winners in this virtual group due to possibly
the addition of other buyers who are in different groups than $u_i$ but with higher reported values 
than that of $u_i$. Even in this case, her dominant strategy is still to report her true value, 
as otherwise she might not be even able to be selected as one of winners in this more competitive environment.  
In addition, Step 3 has no impact on her strategy, as this step only deals with group membership. 

Case 2. Suppose $\hat{b}_i = b_i$ and $u^b_i$ is not chosen for trade by the rule.
Then there are only two possible reasons.  
First, $u^b_i$ cannot pass Step 2 of the algorithm.
This step only uses the vanilla McAfee algorithm, which only uses reported bids and asks
to select users for trade. 
Due to the truthfulness property of McAfee, there is no benefit for $u^b_i$ to cheat about her bid
in a possibly more competitive environment (see the reasoning in Case 1). 
Second, $u^b_i$ can pass Step 2, but cannot pass Step 3. 
Since in Step 3, the algorithm only considers users' group membership information 
in selecting winning users, thus, misreporting bids will not have any impact. 

Secondly, Virtual Market rule is no-deficit or budget balance. In Step 2, it uses McAfee auction 
(see \ref{sec_mcafee} and \cite{mcafee1992dominant}),
in which the total amount of determined payments that will be collected from buyers is always no less 
than the total amount of determined payments that will be sent to sellers. 
In Step 3, the Group Matching algorithm returns the exact same number of buyers and sellers,
and since the McAfee auction in Step 2 determines that all winning buyers pay the same amount of 
payment to the auctioneer and all winning sellers receive the same amount of payment 
from the auctioneer, thus, there is no deficit.   

Thirdly, Virtual Market rule is feasible, as the McAfee auction in Step 2 returns an exact same number of winning
buyers as that of the winning sellers, and the Group Matching algorithm in Step 3 guarantees that 
each final winning buyer is always paired with one and only one final winning seller in the same group, and vice versa.

Finally, Virtual Market rule is individual-rational, as the payment sent by a final winning buyer 
to the auctioneer is always no more than her true value, if she reports her true value in the McAfee auction 
in Step 2. This is guaranteed in a McAfee auction, see Section \ref{sec_mcafee}.
Note that the Group Matching in Step 3 has no impact on payment determination. 
Thus, a final winning buyer should expect non-negative utility if reporting true bid. 
Following a similar reasoning, Virtual Market rule can also guarantee non-negative utility for a seller if she reports her true ask. 
Recall the utility definition is given in Section \ref{sec_cru}.

\end{proof}

\end{document}